\documentclass[12pt]{article}
\usepackage{amsmath}
\usepackage{amssymb}
\begin{document}
\begin{center} 
{\bf{{BRST and Related Superfield Approach to a Few  Interesting  Models of 
Point Particles and a Model of Bosonic String: A Brief  Review}}}  

\vskip 2.8cm

{\sf A. K. Rao$^{(a)}$, B. Chauhan$^{(a)}$, R. P. Malik$^{(a,b)}$ }\\
$^{(a)}$ {\it Department of Physics, Institute of Science,}\\
{\it Banaras Hindu University, Varanasi - 221 005, (U.P.), India}\\

\vskip 0.1cm

$^{(b)}$ {\it DST Centre for Interdisciplinary Mathematical Sciences,}\\
{\it Institute of Science, Banaras Hindu University, Varanasi - 221 005, India}\\
{\small {\sf {e-mails: amit.akrao@gmail.com; bchauhan501@gmail.com;  rpmalik1995@gmail.com}}}
\end{center}

\vskip 1.9 cm

\noindent
{\bf Abstract:}
In this brief  review-cum-research article, we discuss a few key aspects of
the off-shell nilpotent (anti-)BRST symmetry transformations,  Curci-Ferrari (CF) type restriction(s), coupled Lagrangians/Lagrangian densities, etc., for the 1D diffeomorphism (i.e. reparameterization) invariant models of $(i)$ a non-relativistic and non-SUSY free particle, $(ii)$ a scalar (i.e. non-SUSY) relativistic  free particle, $(iii)$ a spinning (i.e. SUSY) relativistic free particle, and $(iv)$ a 2D 
diffeomorphism invariant model of a specific bosonic string theory within the framework of BRST and related supervariable/superfield
approach. We take up a {\it new} 1D diffeomorphism  invariant model of an {\it interacting} scalar relativistic particle 
with the electromagnetic field. The {\it latter} is treated as a constant  background variable and is,
therefore,  independent of the evolution  parameter. We show that the universal nature of the CF-type restriction is
maintained in the case of this {\it new} model, too. 
We exploit the modified Bonora-Tonin supervariable/superfield approach (MBTSA) to BRST formalism
in the context of the 1D and 2D diffeomorphism invariant theories to prove the {\it universal} nature of the CF-type restriction(s). 
We further demonstrate that the 1D diffeomorphism invariant models of {\it all} kinds of particles are 
described by the {\it singular}  Lagrangians. This singular nature has, to the best of our knowledge, {\it not} been explicitly 
shown {\it elsewhere}. Thus, these 1D models {\it also} respect the gauge symmetry transformations which  
 are generated by  the  first-class constraints that exist on them. 
The equivalence of the 1D diffeomorphism and gauge symmetries  is {\it also} established under specific conditions.

\vskip 0.7 cm
\noindent
PACS numbers: 11.15.-q; 12.20.-m; 11.30.Pb.; 02.20.+b; 04.60.Cf.; 11.25.Sq; 11.30.-j \\

\vskip 0.3 cm
\noindent
{\it {Keywords}}: Diffeomorphism invariance in 1D and 2D; free (non-)relativistic and (non-)SUSY particles; interacting scalar relativistic particle
with the background electromagnetic field; a model of bosonic string theory; 
BRST formalism; (anti-)BRST symmetries; supervariable/superfield approach; Curci-Ferrari (CF) type restriction(s)

\newpage
\section {Introduction}

A comprehensive theoretical study of the one (0+1)-dimensional (1D) diffeomorphism (i.e. reparameterization) invariant
relativistic point particles of various kinds (e.g. a free scalar, a free spinning,  an interacting scalar, etc.) has been at the heart of the
modern developments in the realm of the two (1+1)-dimensional (2D) diffeomorphism invariant models of (super)string theories
(see, e.g. [1-5] and reference therein) which are at the forefront areas of research 
activities\footnote{The superstring theories are (i) the most promising candidates for the unification of 
{\it all} the fundamental interactions of nature, (ii) the ones that go {\it beyond} the standard model of particle physics
 which is an incomplete theory due to the conclusive experimental evidence of the {\it mass} of the neutrino, and 
 (iii) the theoretical frameworks which have the  capabilities to provide the theory of quantum gravity 
 (see, e.g. [1-5]). These superstring theories have also led to many other areas of research in mathematics and THEP. } in the domain  of  
theoretical high energy physics (THEP).
Diffeomorphism symmetry invariance {\it also}  plays a decisive role in the description of  the gravitational theories. 
One of the key objectives  of our present endeavor is to  apply the theoretical sophistications of the
Becchi-Rouet-Stora-Tyutin (BRST) and related supervariable approach in the contexts of the 1D diffeomorphism 
(i.e. reparameterization) invariant models of (i) a free non-SUSY non-relativistic particle, (ii) a free non-SUSY
scalar relativistic particle, (iii) a free spinning (i.e. SUSY) relativistic particle, and (iv) an {\it interacting} scalar relativistic particle in interaction with an electromagnetic field that remains in the background. We {\it also} exploit the beauty of the 
BRST and related superfield approach in the context of a 2D diffeomorphism invariant model of a bosonic string  which was 
proposed by Kato and Ogawa [6]. We discuss the off-shell nilpotent and absolutely anticommuting (anti-)BRST symmetry 
transformations with a bit of  emphasis on the existence of the (anti-)BRST invariant Curci-Ferrari (CF) type restriction(s)
for the models under consideration.

Against the backdrop of the above paragraph, we would like to lay emphasis on the supervariable/superfield 
approach to BRST formalism which has been proposed  in our recent works [7, 8] where the 
infinitesimal D-dimensional  diffeomorphism symmetry transformations 
have been  taken into account. We have christened this approach as the {\it modified} Bonora-Tonin supervariable/superfield approach
(MBTSA) to BRST formalism which is the generalization of the earlier works  by Bonora, Pasti and Tonin [9-11] that have  been 
applied to the D-dimensional non-Abelian 1-form  gauge theory to obtain the (anti-) BRST symmetry transformations and Curci-Ferrari condition [12]
by exploiting the geometrical horizontality condition where the exterior derivative of differential geometry plays an important role.
 The theoretical strength  of the supervariable/superfield approach
to BRST formalism [9-11, 13-17] lies in the fact that the appropriate (anti-)BRST symmetry transformations are obtained {\it without} any knowledge 
of the full (anti-)BRST invariant Lagrangians/Lagrangian densities for a given $p$-form ($ p = 1, 2, 3,...$) gauge theory.  
Once the (anti-)BRST symmetries are known, there is a standard technique of BRST formalism [18-21] which enables us to obtain the 
gauge-fixing and Faddeev-Popov (FP) ghost terms of the (anti-)BRST invariant  Lagrangians/Lagrangian densities.
This is how the {\it full} (anti-)BRST invariant Lagrangians/Lagrangian densities  are precisely obtained/constructed.

We would like to dwell a bit on the physical meaning of the horzontality condition (HC) that was first used in [9] for the derivation
of the (anti-)BRST symmetries as well as the CF-condition [12] for the D-dimensional {\it ordinary} non-Abelian 1-form gauge theory
(without any interaction with matter fields) by using the superfield approach to BRST formalism. In this approach, the D-dimensional ordinary spacetime
(parametrized by the coordinates $x^\mu$) 
is generalized onto the (D, 2) dimensional superspace characterized by the superspace coordinates $Z^M = (x^\mu, \theta, \bar\theta)$
where a pair of Grassmannian variables satisfy: $\theta^2 = \bar \theta^2 = 0, \; \theta \,\bar\theta + \bar \theta \,\theta = 0$
and $\mu = 0, 1, 2...D-1$. The field-strength tensor $F_{\mu\nu}$ of the ordinary non-Abelian theory is generalized
onto the (D, 2)-dimensional supermanifold as $F_{\mu\nu} \to \tilde F_{MN} \equiv (F_{\mu\nu}, F_{\mu\theta}, F_{\mu\bar \theta}, F_{\theta\theta},  
F_{\theta\bar\theta}, F_{\bar\theta \bar\theta})$ where $\tilde F_{MN} $ is the super field-strength tensor that encompasses in its folds
the ordinary field-strength tensor $ F_{\mu\nu}$ and the {\it other} Grassmannian components. The celebrated HC requires that the ordinary 
field-strength tensor $ F_{\mu\nu}$ should {\it not} be affected by the presence of the Grassmannian variables which are {\it only} the mathematical artifacts
of superspace (and they have nothing to do with the physical quantities). In more precise words, due the HC, we set 
{\it all} the Grassmannian components of the super field-strength tensor equal to zero 
(i.e. $F_{\mu\theta}= F_{\mu\bar \theta}= F_{\theta\theta}= F_{\theta\bar\theta}= F_{\bar\theta \bar\theta} = 0$). This leads to the derivation of (i) the (anti-)BRST symmetry transformations for
the gauge and (anti-)ghost fields, and (ii) the (anti-)BRST invariant CF-condition [12]. In the Abelian limit, this HC becomes
more transparent in the sense that it requires that the electric and magnetic fields (which are the components of the Abelian field-strength tensor) 
are unaffected by the presence of the Grassmannian variables (which are only the {\it mathematical} tools of superspace formulation). Similar kinds of
superspace restrictions have {\it also} been used in the context of 
our 1D and 2D diffeomporphism invariant theories (cf. Secs. 6,8).

The existence of the CF-type restriction(s), in the realm of BRST approach to gauge/diffeomorphism invariant 
theories, is as fundamental at the {\it quantum} level as the existence of the first-class constraints (see, e.g. [22, 23] for details) for the 
definition of a {\it classical} gauge theory. In fact, the first-class constraints generate the {\it classical}
infinitesimal and continuous {\it local} gauge symmetry transformations  (see, e.g. [22-27] for details) for the systems that belong to the specific class 
of theories which are known as the  gauge theories (cf. Appendix D). 
We pinpoint the CF-type restriction(s) in the context of the (anti-)BRST
symmetry transformations corresponding to (i) the local gauge transformations for the 1D diffeomorphism invariant models  of
the (non-)SUSY, (non-)interacting and (non-)relativistic particles, and (ii) the local diffeomorphism  
symmetry transformations of the 1D and 2D models of the point particles and a specific model of the bosonic string theory [6], respectively. 
We exploit the mathematical sophistication of the MBTSA to BRST formalism [8] to derive 
(cf. Secs. 6 and 8 for details) the (anti-)BRST invariant  
CF-type restriction(s) in the context of the 1D 
and 2D diffeomorphism invariant models under consideration in our review-cum-research article.

In our present endeavor, we have focused on the 1D and 2D diffeomorphism invariant theories and derived 
(i) the proper (i.e. off-shell nilpotent and absolutely anticommuting) (anti-)BRST symmetry transformations, and (ii) 
the {\it universal} (anti-)BRST invariant CF-type restriction(s) by exploiting the theoretical potential of the MBTSA to BRST formalism.
The theoretical arsenal of the MBTSA enables us to obtain the proper (anti-)BRST symmetry transformations for {\it only} a specific 
kinds of variables/fields and the (anti-)BRST invariant CF-type restriction(s). However, in our present endeavor, 
we have shown that the (anti-)BRST symmetry transformations for the {\it rest} of the variables
can be computed by exploiting the key 
properties of (i) the nilpotency, (ii) the absolute anticommutativity, and (iii) the (anti-)BRST invariance 
of the CF-type restriction(s). In this context, we have derived the  {\it quantum} (anti-)BRST symmetry transformations for the 1D diffeomorphism 
invariant model of an {\it interacting} scalar relativistic particle (cf. Appendix C below).
One of the key highlights of our present endeavor  is the derivation of the {\it universal} CF-type restriction(s) for the 
1D and 2D diffeomorphism invariant theories which are nothing but the {\it special} cases of the D-dimensional (anti-)BRST invariant CF-type 
restrictions that have been derived in our earlier works [7, 8] on the D-dimensional diffeomorphism invariant theory.  
As we have pointed out earlier, the existence of the (anti-)BRST invariant
CF-type restriction(s) for the {\it quantum} (anti-)BRST invariant theories is as fundamental as the existence is of the first-class 
constraints for the {\it classical} gauge theories {\it because} the CF-type restrictions are (i) connected
 with the geometrical objects called gerbes (see, e.g. [28, 29]), (ii) responsible for the 
validity of the  absolute anticommutativity property of the (anti-)BRST symmetries, and 
(iii) the reasons behind the existence of a set of coupled (but equivalent) (anti-)BRST invariant 
Lagrangians/Lagrangian densities for an (anti-)BRST invariant theory. 
In the case of an Abelian 1-form gauge theory, the CF-type restriction is 
{\it trivial}\footnote{The other such interesting physical systems are the gauge theories of (i) the 
free non-relativistic particle (cf. Sec. 2), (ii) the free scalar relativistic particle (cf. Sec. 3), and (iii) 
the interacting scalar relativistic particle with the {\it background} electromagnetic fields (cf. Sec. 5).
It is interesting to point out that {\it all} these systems are described by a {\it single} (anti-)BRST invariant Lagrangian
(cf. Secs. 2, 3 and 5 for details).} and, hence, only a {\it single} 
(anti-)BRST invariant Lagrangian density exists. Physically, the geometrical objects gerbes provide the 
(anti-)BRST invariant relationship(s) among different kinds of basic and auxiliary fields which possess
the {\it same} ghost number. For instance, in our present endeavor, we have such kinds of relationships [cf. Eqs. (23),(24),(40),(106)]
for the 1D and 2D gauge and diffeomorphism invariant theories. In most of the cases, the 
geometrical objects gerbes connect the Nakanishi-Lautrup auxiliary fields/variables
with the (anti-)ghost fields/variables of our theories.

Our present review-cum-research  article is essential on the following counts. First of all, we have recently written a set of papers (see, e.g. [30-34])
where we have discussed the 1D and 2D diffeomorphism symmetries at the {\it classical} level and their counterparts at the 
{\it quantum} level within the framework of BRST formalism. Thus, it is interesting to collate  all these results {\it together} 
and present them in a unified and {\it coherent} fashion at one place. Second, we have proven the {\it universal} nature [8] of the CF-type restriction(s) for the 1D 
and 2D diffeomorphism invariant theories of the non-interacting, (non-)SUSY and (non-)relativistic particles and a specific model of the bosonic string theory [6].
It is,  therefore, very interesting to show their existence in the {\it ordinary} space as well as in the {\it superspace} 
by using the MBTSA to BRST formalism. 
Third, all the models of 1D diffeomorphism invariant theories of (non-)relativistic, 
(non-)SUSY and (non-)interacting particles are endowed with the {\it singular} Lagrangians. 
We have demonstrated the singular nature by {\it explicit} computations which, to  the best of our knowledge, we have {\it not} seen elsewhere. 
Finally, we have {\it not} discussed in our previous works [30-32]  the 1D diffeomorphism invariant  {\it interacting}
scalar relativistic particle where the interacting electromagnetic field is constrained to remain in the background. It is a challenging 
problem for us to show that (i) the  form of  the (anti-)BRST invariant  CF-type restriction, 
and (ii) the form of the gauge-fixing and FP-ghost terms 
remain the {\it same} as in the cases of the 1D diffeomorphism invariant  {\it free} (non-)SUSY 
and (non-)relativistic particles [30-32]. We have accomplished  this goal, too, in our present work (cf. Sec. 5 for details)
which establishes the universality of some of the key features of the 1D diffeomorphism invariant theories of different varieties.

The theoretical contents of our present endeavor are organized as follows. In Sec. 2, we discuss the infinitesimal, continuous and {\it classical} gauge 
and reparameterization (i.e. 1D diffeomorphism) symmetry transformations and their {\it quantum} (anti-)BRST counterparts 
for the free non-SUSY non-relativistic particle. Our Sec. 3 covers the discussions on the free scalar (i.e. non-SUSY)
relativistic particle and its gauge as well as  reparameterization  symmetries {\it plus}  
their generalizations to the (anti-)BRST symmetries. In Sec. 4, we devote time on the discussion of the {\it classical} 1D 
diffeomorphism (i.e. reparameterization) as well as the (super)gauge symmetry transformations and their generalizations 
to the level of {\it quantum} (anti-)BRST symmetry transformations for the massive spinning (i.e. SUSY) relativistic particle. 
Our Sec. 5 deals with the discussions on an {\it interacting} 
scalar relativistic particle with a constant background electromagnetic field and its {\it classical} infinitesimal 
gauge as well as reparameterization symmetry transformations (plus their {\it quantum} counterparts as the off-shell 
 nilpotent (anti-)BRST symmetry transformations). 
In Sec. 6, we focus on the derivation of the universal Curci-Ferrari (CF) type restriction by exploiting the
 MBTSA to BRST formalism for the cases of {\it all} the above (non-)relativistic, 
(non-)SUSY and (non-) interacting  particles  which respect the infinitesimal and continuous {\it classical} gauge as well as 1D diffeomorphism 
 symmetry transformations. Our Sec. 7 deals with the model of a bosonic string 
theory (proposed by Kato and Ogawa [6]). We quote the infinitesimal and continuous classical 2D diffeomorphism 
symmetry transformations  [33] and their quantum  counterparts (anti-)BRST transformations for the appropriate Lagrangian 
densities [33, 34]. We devote  time, in Sec. 8, on the derivation of the 2D version of the {\it universal} CF-type restrictions
for {\it this} 2D diffeomorphism invariant bosonic string theory by using the MBTSA to BRST formalism [8].
Finally, in Sec. 9, we make some concluding remarks and point out the future perspective  of our present endeavor on the 
diffeomorphism invariant theories of physical interest (e.g. gravitational theories in the physical 4D spacetime).

In our Appendices  A, B, C and D, we discuss some explicit computations that are relevant 
to the main body of the text of our present review-cum-research  article.

\vskip 0.2cm

\noindent
{\it Convention and Notations}:
For all the 1D and 2D diffeomorphism invariant models of our discussions, we have chosen the target space 
to be a D-dimensional Minkowski  spacetime manifold with a {\it flat} metric tensor $\eta_{\mu\nu} =$ diag  $(+1, -1, -1,...)$
so that the dot product between two non-null vectors $A^\mu$ and $B^\mu$ is defined as: $A\cdot  B = \eta_{\mu\nu}\, A^\mu \, B^\nu \equiv 
\eta^{\mu\nu}\, A_\mu \, B_\nu = \partial_0 B_0 - \partial_i B_i$ where the Greek indices 
$\mu, \nu, \lambda,...  = 0, 1, 2,..., D - 1$, the Latin indices $i, j, k,... = 1, 2, 3,...,D - 1$ 
and Einstein's summation convention is  always followed. 
As a consequence, we have {\it also} used the notation: $A^2 = A \cdot A = A_\mu\, A^\mu \,  \equiv \, A_0\, A_0 - A_i\, A_i $ in our present endeavor. 
We have adopted the convention 
of the left derivative w.r.t. {\it all} the fermionic variables/fields and chosen the notations $s_{ab}$ and $s_b$ for the infinitesimal,
and off-shell nilpotent  (anti-)BRST transformations [$s_{(a)b}$], respectively.  
We always follow the notations  $L_{\bar b}$ and  $L_b$ for the coupled (but equivalent) (anti-)BRST invariant Lagrangians, respectively,
in the cases of {\it all} kinds of particles where, at the classical level, we have {\it also} denoted  
the original, first-order and second-order Lagrangians by $L_0, \, L_f$ and $L_s$, respectively. 
Furthermore,  the symbols for the (anti-)ghost variables, corresponding to the {\it classical}  gauge and 1D diffeomorphism  symmetries, are taken to be
$(\bar c)c$ and $(\bar C)C$, respectively. In addition,  the pairs of Nakanishi-Lautrup auxiliary variables/fields for the gauge and 
diffeomorphism invariant theories have been chosen to be $(b, \, \bar b)$ and $(B, \, \bar B)$, respectively. 
We have also used the overdot on the generic variable as: $\dot \Phi = (d\, \Phi / d\, \tau)$ in our text
which denotes the first-order ``time" derivative of the generic variable ($\Phi $).


\section{Gauge and 1D Diffeomorphism Invariant Model of a Non-Relativistic Particle: (Anti-)BRST Symmetries and CF-Type Restriction}

We begin with the gauge and reparameterization non-invariant model of a non-relativistic (NR) particle where the {\it original}   Lagrangian
$L_0^{(NR)}  = \frac {1}{2}\, m\, \dot x^2$ (with $\dot x = dx/dt, \; m = $ mass, $t = $ evolution parameter)
 has been generalized to a different form (i.e. $L_0$) in which $x$ and $t$
both have been treated on equal footing in the configuration space  and they have been assumed to be a  function of
another evolution parameter $\tau$. In other words, we have now {\it two} ``generalized" velocities:  
 $\dot x = d x / d \tau$ and $\dot t = d t/ d \tau$ in the theory. This has been done because, in the 
 literature (see, e.g. [35-37]), the ``space" and ``time" (i.e. $x$ and $t$) have been treated as observable operators at the 
quantum level. In our earlier works, we have derived  {\it three} equivalent reparameterization-invariant   Lagrangians for this 
generalized version of a free non-relativistic particle as (see, e.g. [30, 36, 37] for details)
\begin{eqnarray}
&&L_0(x,\,\dot x, \, t, \dot t) = \frac{m\,{\dot x}^2}{2\,\dot t},  \nonumber \\
&&L_f(x,\,\dot x, \, t, \dot t,\,p_x,\,p_t) = p_x\,\dot x + p_t\,\dot t - \frac{1}{2}\,E\,(p_x^2 + 2\,m\,p_t), \nonumber \\
&&L_s(x,\,\dot x, \, t, \dot t) = \frac{{\dot x}^2}{2\,E} + \frac{m\,{\dot x}^2}{2\,\dot t}\Big[\frac{E\,m}{\dot t} - 1\Big],
\end{eqnarray}
where $E(\tau)$ is a Lagrange multiplier variable that is needed so that the constraint: $p_x^2 + 2\,m\,p_t \approx 0$ can be incorporated 
 into the first-order Lagrangian $L_f(x,\,\dot x, \, t, \dot t,\,p_x,\,p_t)$. We have 
$p_x = ({\partial \,L}/{\partial\, \dot x})$ and  $p_t = ({\partial \,L}/{\partial\, \dot t})$ as canonically conjugate momenta variables w.r.t. $x$ and $t$
where $L$ stands  for  $L_0$,  $L_f$ and $L_s$.
It can be {\it trivially} checked  that we have $\dot p_x = 0, \; \dot p_t = 0$ as the Euler-Lagrange  equations 
of motion (EL-EoMs) from $L_f$ which shows that the {\it free}  nature of a non-relativistic 
particle is retained in the generalized version of our theory, too. 
We point out that  $L_0 (x,\,\dot x, \, t, \dot t)$ is the generalized version of the 
original Lagrangian: $L_0 ^ {(NR)} = \frac {1}{2}\,m\,\dot x^2$. The {\it latter} Lagrangian  has {\it no} reparameterization (i.e. 1D diffeomorphism)
and gauge symmetry  transformations because it is {\it not} a singular Lagrangian and it has {\it no} diffeomorphism symmetry transformations 
in terms of its evolution parameter $t$.

A few remarks, at this juncture, are pertinent here. First, the original Lagrangian $L_0^ {(NR)}  (x, \dot x)= \frac {1}{2}\,m\,\dot x^2$, where 
$t$ is the evolution parameter,  has {\it no} constraints and it does {\it not} respect the reparameterization (i.e. 1D diffeomorphism) 
symmetry transformation in terms of its evolution parameter $t$. In other words, the 1D diffeomorphism transformations: 
$t \longrightarrow \ t^\prime = f (t) \equiv t - \epsilon (t) $ where $f(t)$ is any physically well-defined function of $t$ 
(which is finite at the origin and  vanishes off as $t \longrightarrow \pm \,\infty$ and $\epsilon (t)$ is an 
infinitesimal transformation parameter),
is {\it not} a symmetry transformation for the  action integral $S = \int_{- \, \infty}^{+ \, \infty} d t\, \, L_0 ^ {(NR)} (x, \dot x)$ defined in terms of 
$L_0 ^ {(NR)} (x, \dot x)$. 
Second,  the Lagrangians   $L_0 (x,\,\dot x, \, t, \dot t),\; L_f(x,\,\dot x, \, t, \dot t,\,p_x,\,p_t)$ and $L_s(x,\,\dot x, \, t, \dot t) $
 are endowed with a reparameterization symmetry transformations: $\tau \longrightarrow \ \tau^\prime = f (\tau) \equiv \tau - \varepsilon (\tau) $
where $\varepsilon  (\tau)$ is an infinitesimal parameter and $f (\tau)$ is a physically well-defined function of $\tau$.
 Third, we note that $L_0$ has no {\it massless} limit and it has a variable in the denominator 
(as is  the case with $L_s(x,\,\dot x, \, t, \dot t)$). Fourth, the {\it massless} limits of $L_f$ and $L_s$ are possible.
However, the first-order  Lagrangian is having superiority over  $L_s$ in  the sense that  it has $(i)$ more 
dynamical variables at our disposal,  and $(ii)$  {\it no} variables in the denominator. Fifth, the form of
 $L_s (x, \dot x, t, \dot t)$ is {\it correct} because the second-order 
terms in $\dot x$ and $\dot t$ exist in it.  Finally, we shall be dealing 
with the first-order Lagrangian ($L_f$) because it has {\it no} variable in the denominator and {\it easier} to  deal with theoretically because there are 
more variables in it compared to $L_0$ and $L_s$.

We focus, first of all, on $L_0 (x,\,\dot x, \, t, \dot t)$ and note that it is  a {\it singular} Lagrangian because of the 
fact that the determinant of the following $2 \times 2 $-matrix (see, e.g. [24, 23])
\begin{eqnarray}
W \; = \;
\renewcommand\arraystretch{2}\large{
\begin{pmatrix}
\frac{{\partial^2 \,L_0}}{{\partial\, \dot x \, \partial\, \dot x}} \;  & \; \frac{{\partial^2 \,L_0}}{{\partial\, \dot x \, \partial\, \dot t}}
 \\ \frac{{\partial^2 \,L_0}}{{\partial\, \dot t \, \partial\, \dot x}}  \;  & \;\frac{{\partial^2 \,L_0}}{{\partial\, \dot t \, \partial\, \dot t}}
\end{pmatrix}
\quad \equiv \quad 
\renewcommand\arraystretch{2}
\begin{pmatrix}
\frac{m}{\dot t}  &  -\, \frac{m\, \dot x}{\dot t^2}  \\
-\, \frac{m\, \dot x}{\dot t^2} &  \frac{m\, \dot x^2}{\dot t^3}
\end{pmatrix}},
\end{eqnarray}
turns out to be zero (i.e. det $W = 0$). This is why there is a constraint ($p_x^2 + 2\,m\,p_t \approx 0$) on this theory where 
we have the following expressions for the conjugate momenta: 
\begin{eqnarray}
p_x = \frac{{\partial \,L_0}}{{\partial\, \dot x}}   \equiv  \frac{m\, \dot x}{\dot t}, \qquad
p_t = \frac{{\partial \,L_0}}{{\partial\, \dot t}}     \equiv -\, \frac{m\, \dot x^2}{2\, \dot t^2}.
\end{eqnarray}
This constraint (i.e. $p_x^2 + 2\,m\,p_t \approx 0$) has been incorporated into the first-order
 Lagrangian $L_f(x,\,\dot x, \, t, \dot t,\,p_x,\,p_t)$ through the 
Lagrange multiplier variable $E\,(\tau)$ because the EL-EoM w.r.t. $E\,(\tau)$, derived  from  $L_f$, yields it.
To be more precise, we have {\it two} constraints on our theory (described by $L_f$) where $\Pi_E \approx 0$ is the primary constraint 
and $p_x^2 + 2\,m\,p_t \approx 0$ is the secondary constraint. Here $\Pi_E$ is the conjugate momentum w.r.t. the Lagrangian multiplier  
variable $E(\tau)$. It is obvious that these  two constraints are first-class in the terminology of Dirac's prescription for the 
classification of constraints [22, 23]. In our earlier work [37], we have shown that these constraints generate the following {\it classical}, 
infinitesimal and continuous gauge transformations $(\delta_g)$ for the first-order Lagrangian $L_f$
\begin{eqnarray}
\delta_g\, x  = \xi\, p_x, \qquad \delta_g \,t  = \xi \, m, \qquad \delta_g \,E = \dot \xi, \qquad \delta_g \,p_x  = 0 \, \qquad \delta_g p_t = 0,
\end{eqnarray}
because we observe that the following is true, namely; 
\begin{eqnarray}
\delta_g\, L_f = \frac {d}{d\tau} \, \Big [\frac {1}{2}\, \xi\, p_x^2\Big].
\end{eqnarray}
As a consequence, the action integral $S = \int_{- \, \infty}^{+ \, \infty} \, d\tau\, L_f$ remains invariant 
(i.e. $\delta_g\, S = 0$) for the physically well-defined infinitesimal gauge symmetry parameter $\xi (\tau)$ and the 
conjugate momentum $p_x (\tau)$ which vanish off as $\tau \longrightarrow \pm\, \infty$.
The full set of off-shell nilpotent $[s_{(a)b}^2 = 0]$ and absolutely anticommuting ($s_b\,s_{ab} + s_{ab}\, s_b = 0$)
(anti-)BRST symmetry transformations $[s_{(a)b}]$ for our system  are as follows (see, e.g. [30, 37] for details):  
\begin{eqnarray}
 && s_{ab} \, x = \bar c\, p_x, \quad s_{ab}\, p_x  = 0,  \quad s_{ab} \,c  = -\, i\,b, \quad s_{ab} \, \bar c = 0,
 \quad s_{ab} \, t  = \bar c\, m,\nonumber\\
 && s_{ab}\,  b = 0, \quad s_{ab} E = \dot {\bar c}, \quad s_{b} \,x = c\, p_x, \quad s_{b} \,p_x  
= 0, \quad s_{b}\, c  = 0, \quad s_{ab}\, \bar c = i\, b,\nonumber\\
 && s_{b} \, t  =  c\, m, \quad s_{b}\,  b = 0, \quad s_b E = \dot c.
\end{eqnarray}
Under the above nilpotent (anti-)BRST symmetry transformations, the following (anti-) BRST invariant Lagrangian 
with the gauge-fixing term along with the Nakanishi-Lautrup type auxiliary variable $b$
and the fermionic ($c^2 = \bar c^2, \; c\, \bar c + \bar c\, c = 0$) Faddeev-Popov (anti-) ghost  variables  $(\bar c)\, c$, namely;
\begin{eqnarray}
L_b &=& L_f + b\,\dot E + \frac{b^2}{2} - i\,\dot{\bar c}\,\dot c, \nonumber\\
&\equiv& p_x\,\dot x + p_t\,\dot t - \frac{1}{2}\,E\,(p_x^2 + 2\,m\,p_t) + b\,\dot E + \frac{b^2}{2} - i\,\dot{\bar c}\,\dot c, 
\end{eqnarray}
transforms to the total time derivatives as follows:
\begin{eqnarray}
s_{ab}\,L_b = \frac{d}{d\,\tau}\,\Big[\frac{\bar c}{2}\,p_x^2 + b\,\dot{\bar c}\,\Big], \qquad  
s_b\,L_b = \frac{d}{d\,\tau}\,\Big[\frac{c}{2}\,p_x^2 + b\,\dot c\,\Big].
\end{eqnarray}
Hence, the action integral $S = \int_{- \, \infty}^{+ \, \infty} \, d\tau\, L_b$ remains 
invariant (i.e. $s_{(a)b\, S = 0}$) under the off-shell nilpotent  (anti-)BRST symmetry transformations.

In addition to the above classical, infinitesimal and continuous gauge symmetry transformations, the Lagrangian $L_f$
respects the following classical, infinitesimal and continuous 1D diffeomorphism (i.e. reparameterization) symmetry transformations 
$(\delta_r)$
\begin{eqnarray}
\delta_r \,x = \varepsilon\,\dot x, \quad\quad \delta_r \,p_x = \varepsilon\,{\dot p}_x, \quad\quad 
\delta_r \,t = \varepsilon\,\dot t, \quad\quad \delta_r \,p_t = \varepsilon\,{\dot p}_t,   \quad\quad 
\delta_r \,E = \frac{d}{d\,\tau}\,(\varepsilon\,E), \quad\qquad 
\end{eqnarray}
because we observe  that the following is true: 
\begin{eqnarray}
\delta_r \,L_f = \frac{d}{d\,\tau}\,\Big [ \varepsilon\,L_f \Big] 
 \quad \Longrightarrow \quad  \delta_r \, S = \int_{- \, \infty}^{+ \, \infty} \, d\tau\, \delta_r \, L_f \quad \Longrightarrow  \quad 0.
\end{eqnarray}
At this juncture, it is very interesting to mention that the gauge symmetry transformations (4) can be derived from the 1D 
diffeomorphism symmetry transformations (9) if we $(i)$ identify $\xi(\tau) = \varepsilon (\tau)\, E(\tau)$,  and $(ii)$
use the equations of motion: $\dot p_x = \dot p_t = 0, \, \dot x = E\, p_x,\, \dot t = E\, m$. 
The full set of (anti-)BRST symmetry transformations $[s_{(a)b}]$, corresponding to the above {\it classical} 
reparameterization symmetry transformations (9), are (see, e.g. [30])
\begin{eqnarray}
&&s_{ab}\,x = \bar C\,\dot x,\quad s_{ab}\,p_x = \bar C\,\dot{p_x},\quad s_{ab}\,t = \bar C\,\dot t,\quad s_{ab}\,p_t = \bar C\; \dot p_t,\quad 
s_{ab}\,E = \frac{d}{d\,\tau}\,(\bar C\, E),\nonumber\\
&&s_{ab}\,C = i\, \bar B,\quad s_{ab}\, \bar C = \bar C\, \dot {\bar C}, \quad s_{ab}\,\bar B = 0, \quad  s_{ab}\, B = \dot {B}\, \bar C - B\, \dot{\bar C},
\nonumber\\
&&s_{b}\,x = C\,\dot x,\quad~ s_{b}\,p_x = C\,\dot{p_x},\quad~ s_{b}\,t = C\,\dot t,\quad~~ s_{b}\,p_t = C\; \dot p_t,\quad 
s_{b}\,E = \frac{d}{d\,\tau}\,( C\, E),\nonumber\\
&& s_{b}\,\bar C = i\, B,\quad s_{b}\, C = C\, \dot C, \quad s_{b}\, B = 0, \quad  s_{b}\, \bar B = \dot {\bar B}\, C - \bar B\, \dot C,
\end{eqnarray}
where $B$ and $\bar B$ are the Nakanishi-Lautrup type auxiliary variables and fermionic 
$(C^2 = 0, \; \bar C^2 = 0, \; C\, \bar C + \bar C\, C = 0)$ (anti-)ghost variables 
$(\bar C)C$ are the generalization of the {\it classical} reparameterization transformations parameter 
$\varepsilon (\tau)$ that is present in the fundamental 1D diffeomorphism symmetry  
transformation: $\tau \longrightarrow \tau^\prime = \tau - \varepsilon (\tau)$.

A few comments, at this stage, are in order. First, we observe that the (anti-)BRST symmetry transformations 
$[s_{(a)b}]$ are off-shell nilpotent $[s_{(a)b}^2 = 0]$ of order two. Second, we note that the absolute 
anticommutativity property (i.e. $\{s_b, \; s_{ab}\} = 0$) is accurately  satisfied for {\it all} the variables 
except $\phi (\tau) = x, p_x, t, p_t$ and $E$. In other words, for the generic variable $\phi (\tau)$ and $E(\tau)$, we 
note the following: 
\begin{eqnarray}
&& \{s_b,\; s_{ab}\}\, \phi =  i\, [B + \bar B + i\, (\bar C \, \dot C - \dot {\bar C}\, C)]\, \dot\phi, \nonumber\\
&& \{s_b,\; s_{ab}\}\, E = \frac {d}{d\tau}\, \Big[i\,\{B + \bar B + i\, (\bar C \, \dot C - \dot {\bar C}\, C)\}\,E\Big]. 
\end{eqnarray}
Thus, it is obvious that the absolute anticommutativity property (i.e. $\{s_b, s_{ab}\} = 0$) is satisfied iff:
 $B + \bar B + i\, (\bar C \, \dot C - \dot {\bar C}\, C) = 0$.
This condition is popularly  known as the CF-type restriction where the auxiliary variables $B, \, \bar B$ and (anti-)ghost variables $(\bar C)C$
are present in a specific fashion. Third, the coupled (but equivalent) Lagrangians that respect the 
above (anti-)BRST symmetry transformations (11) are as follows (see, e.g. [30] for details)   
\begin{eqnarray}
L_{\bar b} &=& L_f  - \bar B\,\Big[E\,\dot E - i\,(2\, {\bar C}\,\dot C +  \dot{\bar C}\,C)\Big]+\frac{{\bar B}^2}{2} \nonumber\\
&&- i\,E\,\dot E\,{\bar C}\,\dot C - \,i\,E^2\,\dot{\bar C}\,\dot C  - \dot{\bar C}\,{\bar C}\,\dot C\,C,\nonumber\\
L_b &=& L_f  + B\,\Big[E\,\dot E -i\, (2\, \dot{\bar C}\, C  + {\bar C}\,\dot C)\Big]+ \frac{B^2}{2} \nonumber\\
&&- i\,E\,\dot E\,\dot{\bar C}\,C - \,i\,E^2\,\dot{\bar C}\,\dot C  - \dot{\bar C}\,{\bar C}\,\dot C\,C, 
\end{eqnarray}
where $L_f$ is the first-order Lagrangian of Eq. (1). It is quite
straightforward to check that we have the following transformations for the Lagrangians under $s_{(a)b}$, namely; 
\begin{eqnarray}
&&s_{ab}\,L_{\bar b} = \frac{d}{d\,\tau}\Big[\bar C\,L_f + {\bar B}^2\,\bar C  
- i\,\bar B\,\dot{\bar C}\,\bar C\,C - E\,\dot E\,\bar B\,\bar C - E^2\,\bar B\,\dot{\bar C}  \Big],
\end{eqnarray}
\begin{eqnarray}
&&s_b\,L_b = \frac{d}{d\,\tau}\Big[C\,L_f + B^2\,C - i\,B\,\bar C\,\dot C\,C + E\,\dot E\,B\,C  + E^2\,B\,\dot C \Big],
\end{eqnarray}
which establishes that there is a {\it prefect} (anti-)BRST symmetry for the action integrals 
$S_1 = \int_{- \, \infty}^{+ \, \infty} \, d\tau\, L_{\bar b}$ and 
$S_2 = \int_{- \, \infty}^{+ \, \infty} \, d\tau\, L_{b}$ for the anti-BRST and BRST symmetries [cf. Eq. (11)], respectively.
If we utilize the theoretical potential of the (anti-)BRST invariant CF-type restriction: 
$B + \bar B + i\, (\bar C \, \dot C - \dot {\bar C}\, C) = 0$, it can be checked that $L_{\bar b}$ and $L_{b}$ {\it also} respect BRST and 
anti-BRST symmetries, respectively (see, e.g. [30] for details). Hence, $L_{ b}$ and  $L_{\bar b}$ are {\it equivalent} from the 
point of view of the (anti-)BRST symmetries.

We end this section with the following remarks. First, the (anti-)BRST invariant CF-type restriction 
(i.e $s_{(a)b\, }[B + \bar B + i\, (\bar C \, \dot C - \dot {\bar C}\, C)] = 0$)  is {\it universal} for the 1D 
diffeomorphism invariant systems that can always be derived from the coupled (but equivalent) Lagrangians. Second, the existence
of the CF-type restriction is responsible for the absolute anticommutativity of the (anti-)BRST symmetry transformations 
[see, e.g.  Eqs. (11) and (12)] . Third, the existence of the coupled (but equivalent) Lagrangians $L_{\bar b}$ and $L_{b}$
owe their origin to the CF-type restriction in a  subtle  way (because the EL-EoMs w.r.t. Nakanishi-Lautrup auxiliary variables produce it).
Fourth, in Sec. 6, we shall exploit the theoretical strength of the MBTSA to BRST formalism [8] to derive the 
CF-type restriction which is {\it true} for any 1D diffeomorphism invariant theory of physical interest. 
Finally, it can be checked (using the (anti-)BRST symmetry transformations [cf. Eq. (11)])
that the CF-type restriction is (anti-)BRST invariant. Hence, it is a {\it physical} restriction on our {\it quantum} theory.


\section{(Anti-)BRST symmetries for a Free Scalar Relativistic Particle: A  Model for the Gauge and 1D Diffeomorphism Invariant Theory}

In this section, we discuss the bare essentials of the {\it classical} gauge, reparameterization (i.e. 1D diffeomorphism) 
and {\it quantum} (anti-)BRST symmetry transformations for the one (0 + 1)-dimensional (1D) model of a free scalar (i.e. non-SUSY) relativistic particle
that is embedded in the D-dimensional target space (which is nothing but the  D-dimensional Minkowski flat spacetime manifold). 
This model is physically important because it is the precursor to the bosonic string theory as its generalization produces the {\it latter}
 (see, e.g. [1-5] for details). This system has been studied within the framework of BRST 
 formalism by others\footnote{In the work by Marnelius [38], a great deal of emphasis is laid on the 
 inner product space while dealing with the free relativistic scalar particle within the realm of BRST approach
to its proper quantization. On the other hand, in [39], a 
 field theoretic approach has been developed and it has been applied to the system of a free relativistic particle that 
was proposed by Brink, Di Vecchia, Howe and Polyakov. We have followed 
the discussion on the system of a free scaler relativistic particle that has been covered in [40].}, too (see, e.g. [38, 39] and references therein).
This system can be described by three equivalent Lagrangians which have been derived earlier as (see, e.g. [40, 31]) 
\begin{eqnarray}
&& L_0 = m\,\sqrt {\dot x^2},\qquad L_f = p_\mu\,\dot x^\mu - \frac {e}{2}\;(p^2 - m^2),\qquad L_s  = \frac {1}{2\,e}\; \dot x^2 + \frac {e}{2}\;m^2,
\end{eqnarray}
where $m$ is the rest mass of the free scalar relativistic particle, $x_\mu$ and $p^\mu$ are the canonically conjugate space and momenta
variables in the target space (with $\mu = 0, 1 , 2,...,D-1$), $e(\tau)$ is the einbein variable  
and $\dot x^\mu  = (dx^\mu/d\tau)$ is the generalized velocity of the free particle where 1D trajectory  is parameterized by the evolution parameter $\tau$
(and this 1D trajectory is embedded in the D-dimensional target spacetime Minkowskian manifold  where  the mass-shell condition: $p^2 - m^2 = 0$ is satisfied).
The original Lagrangian $L_0$ has a square root and it does {\it not} have the massless limit  and second-order Lagrangian $L_s$ has a variable 
in the denominator. On the contrary, the first-order Lagrangian $L_f$ has maximum number of variables and it is {\it not} 
plagued with the kinds of problems that $L_0$ and $L_s$ are endowed with. Thus, we shall focus on the first-order Lagrangian $L_f$
for our further discussions (as it also has a {\it massless} limit).

It is straightforward to note that the following infinitesimal, continuous and {\it classical} gauge symmetry transformations ($\delta_g$)
and reparameterization (i.e. 1D diffeomorphism) symmetry transformations ($\delta_r$), namely; 
\begin{eqnarray}
&& \delta_g\, x_\mu = \xi\,p_\mu, \qquad \delta_g\, p_\mu = 0, \quad \qquad \delta_g e = \dot\xi, \nonumber\\
&& \delta _r\, x_\mu = \varepsilon \,\dot x_\mu, \quad \delta _r\, p_\mu = \varepsilon  \,\dot p_\mu,\quad
 \delta _r\,e = \frac {d}{d\tau}\, (\varepsilon  \,e),
\end{eqnarray}
transform the Lagrangian $L_f$ in the following manner
\begin{eqnarray}
&& \delta _g\, L_f = \frac {d}{d\tau}\;\Big [\frac {1}{2}\; \xi\, (p^2 + m^2) \Big], \qquad \quad \delta _r\, L_f = \frac {d}{d\tau}\,(\varepsilon  \,L_f),
\end{eqnarray}
where $\xi(\tau)$ and $\varepsilon (\tau)$ are the infinitesimal {\it classical} gauge and reparameterization (i.e. 1D diffeomorphism)
symmetry transformation parameters. The latter is present in $\tau \longrightarrow \tau - \varepsilon (\tau)$. 
As a consequence, the action integral $S = \int_{-\,\infty}^{+\, \infty} d\tau\, L_f $ remains invariant 
(i.e. $\delta_g S = \delta_r S = 0$). It is very interesting to point out that the gauge symmetry transformations 
$(\delta_g)$ can be derived from the reparameterization symmetry transformations $(\delta_r)$ 
if we (i) identify $\xi(\tau) = e (\tau) \, \varepsilon (\tau)$, and (ii) use the equations of motion: $\dot p_\mu = 0$ 
and $\dot x_\mu = e\, p_\mu$. 
The  classical {\it gauge}  symmetry transformations $(\delta_g)$ can be elevated to their {\it quantum} counterparts 
(anti-) BRST symmetry transformations $[s_{(a)b}]$ as follows
\begin{eqnarray}
&&s_{ab}\, x_\mu = \bar c \;p_\mu, \; \; s_{ab} \,\bar c = 0, \; \; s_{ab}\, p_\mu = 0, 
 \;\; s_{ab}\, c = - i\, b,\;\; s_{ab}\, b = 0, \;\; s_{ab}\,\; e = \dot {\bar c},\nonumber\\
&&s_b\, x_\mu = c \,  p_\mu, \;\;\;s_b\, c = 0, \;\;\;\; s_b \,p_\mu = 0, 
\;\;\;\; s_b\, \bar c = i\, b, \;\;\;\;\; s_b \,b = 0, \;\;\;\;\; s_b\, \, e = \dot c,  
\end{eqnarray}
where the fermionic  $(c^2 = \bar c^2 = 0, \; c\, \bar c\,  + \bar c\, c = 0)$  
variables are the (anti-)ghost variables $(\bar c)c$ and $b$ is the Nakanishi-Lautrup 
type auxiliary variable. It is straightforward to check that the above (anti-)BRST symmetry transformations [$s_{(a)b}$] are 
$(i)$ off-shell nilpotent [$s_{(a)b}^2 = 0$], and 
$(ii)$ absolutely  anticommuting  (i.e. $s_b\,s_{ab} + s_{ab}\, s_b = 0$) when they are  treated in their operator forms.
We point out  that the CF-type restriction is {\it trivial}  in our present case.
It is interesting to check that the following {\it single} Lagrangian 
\begin{eqnarray}
L_{b} =  L_f + b\; \dot e 
+ \frac{1}{2}\; b^2 - i\; \dot {\bar c}\; \dot c  \quad \equiv \quad p_\mu \; \dot x^\mu - \frac{1}{2}\;e\; (p^2 - m^2) + b\; \dot e 
+ \frac{1}{2}\; b^2 - i\; \dot {\bar c}\; \dot c.
\end{eqnarray}
remains invariant under the above (anti-)BRST symmetry transformations: 
\begin{eqnarray}
s_{ab} L_b = {\displaystyle \frac{d} {d \tau}} \;
\Bigl [ \frac{1}{2}\; \bar c\; (p^2 + m^2) + b\; \dot {\bar c} \Big ], \quad 
&&s_b L_b = {\displaystyle \frac{d} {d \tau}} \;
\Bigl [ \frac{1}{2}\; c\; (p^2 + m^2) + b\; \dot c \Bigr ].
\end{eqnarray}
To be precise, it is the action integral  $S = \int_{-\,\infty}^{+\, \infty} d\tau\, L_b $ that remains invariant  $[s_{(a)b}\, S = 0]$
under the infinitesimal and continuous (anti-)BRST symmetry transformations.

We now concentrate on the {\it quantum} (anti-)BRST symmetry transformations corresponding to the {\it classical}, infinitesimal
and continuous  reparameterization (i.e. 1D diffeomorphism) symmetry transformations which are as follows (see, e.g. [31] for details):
\begin{eqnarray}
&&s_{ab} x_\mu = \bar C\, \dot x_\mu, \,\; s_{ab} p_\mu = \bar C\, \dot p_\mu, \,\;  s_{ab} C = i\,\bar B,
\;  s_{ab} \bar C = \bar C\,\dot {\bar C}, \nonumber\\
&& s_{ab} e = \frac {d}{d\tau}\,(\bar C\,e),\;\;\; s_{ab} \bar B = 0,\quad \;\; s_{ab} B = \dot{B}\,\bar C - B\,\dot {\bar C},\nonumber\\
&&s_b x_\mu = C\, \dot x_\mu, \; \; s_b p_\mu = C\, \dot p_\mu, \;\; s_b C = C\,\dot C,\;\,\; s_b \bar C = i\, B, \nonumber\\
 && s_b e = \frac {d}{d\tau}\,(C\,e),\quad\; s_b B = 0, \quad \; s_b \bar B = \dot{\bar B}\,C - \bar B\,\dot C. 
\end{eqnarray}
The noteworthy points related to  the above symmetries are (i) they are off-shell nilpotent, and (ii) they are absolutely anticommutativity 
provided the CF-type restriction: $B + \bar B + i \,(\bar C \, \dot C - \dot {\bar C} \, C) = 0$ is taken into account. To be precise, we obtain the following:
\begin{eqnarray}
&&\{s_b, s_{ab}\}\, x_\mu = i\, [B + \bar B + i\,(\bar C\, \dot C - \dot {\bar C}\, C)]\;\dot x_\mu = 0,  \quad (\dot x_\mu \ne 0), \nonumber\\
&&\{s_b, s_{ab}\}\, p_\mu = i\, [B + \bar B + i\,(\bar C\, \dot C - \dot {\bar C}\, C)]\;\dot p_\mu = 0, \quad (\dot p_\mu \ne 0),\nonumber\\
&&\{s_b, s_{ab}\}\, e = i\, \frac {d}{d\tau}\Big[\big\{B + \bar B + i\,(\bar C\, \dot C - \dot {\bar C}\, C)\big\}\,e\Big] = 0, \quad (e \ne 0),\nonumber\\
&&\{s_b, s_{ab}\}\, C = 0,\quad\{s_b, s_{ab}\}\, \bar C = 0,\nonumber\\
&&\{s_b, s_{ab}\}\, B = 0, \;\quad\{s_b, s_{ab}\}\, \bar B = 0,
\end{eqnarray}
which demonstrates that the CF-type restriction has to be invoked  for the proof of the absolute anticommutativity property. 
It is very interesting to point out that the above CF-type restriction is (i) the (anti)BRST invariant quantity, namely; 
\begin{eqnarray}
s_{(a)b}\,[B + \bar B + i\,(\bar C\, \dot C - \dot {\bar C}\, C)] = 0,
\end{eqnarray}
provided we take into account the validity of the CF-type restriction. 
Hence, it is a {\it physical} restriction at the {\it quantum} level, and (ii) there is a set of 
the following {\it two} discrete symmetry transformations, namely;
\begin{eqnarray}
B \rightarrow  - \bar B, \quad  \bar B \rightarrow  - B, \quad  \bar C \rightarrow \pm\, i\, C, \quad  C \rightarrow  \pm\, i\, \bar C,  
\end{eqnarray}
under which the CF-type restriction {\it itself} remains invariant. This shows the theoretical beauty and 
 robustness of the CF-type restriction
within the realm of BRST formalism.

A close look at $L_f$ in Eq. (16) demonstrates that the EL-EoM w.r.t. $e(\tau)$ yields:
$p^2 - m^2 = 0$ which is nothing but the mass-shell condition of a massive free relativistic particle. 
In more sophisticated terminology, the mass-shell condition is a constraint on our theory. To corroborate 
this statement, it can be checked that the following is {\it true} for the starting Lagrangian: 
$L_0 = m\,\sqrt {\dot x ^2} \equiv m\, \sqrt {\dot x_\mu\, \dot x^\mu}$ (with $\dot x^\mu  = dx^\mu/d\tau$), namely;
\begin{eqnarray}
W_{\alpha\beta} = \frac  {\partial^2 \, L_0}{\partial \,\dot x^\alpha\,\partial \,\dot x^\beta} = \frac {m}{(\dot x ^2)^{{3}/{2}}}\;
\big[\eta_{\alpha\beta}\;\dot x^2 -  \,\dot x_\alpha\,\dot x_\beta \big],
\end{eqnarray}
whose determinant turns out to be zero. In other words, for the simple case of $D = 2$ where $\alpha, \, \beta = 0, \; 1,$ 
we have the following {\it symmetric} $2 \times 2$ matrix (see, e.g. [24, 23]): 
\begin{eqnarray}
W_{\alpha\beta} \; = \;
\renewcommand\arraystretch{2}\large{
\begin{pmatrix}
W_{00} \;  & \; W_{01} 
 \\ W_{10}  \;  & \;W_{11} 
\end{pmatrix}
\quad \equiv \quad \frac {m}{(\dot x ^2)^{{3}/{2}}}\;
\renewcommand\arraystretch{2}
\begin{pmatrix}
-\, {\dot {\underline x^2}}  \; \; &  \; \; {\dot x_0\, {\dot {\underline {x}}}}  \\
{\dot x_0\, \dot {\underline {x}}} \; \; &  \;\; -\, {\dot x_0^2} 
\end{pmatrix},}
\end{eqnarray} 
 where we have used $\eta_{\alpha\beta} = (+1, \, -\,1)$ and $x_\alpha = (x_0, \, x_1) \equiv (x_0, \, -\, \underline {x})$
 so that $x^2  \equiv  x_\mu\,x^\mu = x_0^2 - \underline {x}^2$ [because $x^\mu = (x^0, \, x^1) = (x_0, \, + \underline {x})$].
We have chosen $x^\mu = (x^0, x^1) = (x_0, \underline {x})$ and $x_\mu  = ({x_{0}, x_1}) = (x_0, -\underline{x})$
so that there should be {\it no} confusion between $x^2 = x_\mu\,x^\mu = x_0^2 - \underline {x}^2$ and $\underline {x}^2$
(which is the square of the space component  $\underline {x}$ {\it only}). 
 A close look at the above matrix [cf. Eq. (27)] shows that det $W_{\alpha\beta} = 0$ and, hence, the original Lagrangian, with the square root 
(i.e. $ L_0 = m\,\sqrt {\dot x ^2} $),  is a {\it singular} Lagrangian\footnote {We have {\it not} seen, to the best of our knowledge, 
such kind of explicit verification of the {\it singular} nature of the  starting Lagrangian $L_0$ {\it elsewhere.}}
in the 2D of target space (see, e.g., our Appendix A for $D = 3$ where 
$\mu, \;  \nu,...  = 0, \, 1, \, 2$ ).
Hence, there is a constraint: $p^2 - m^2 = 0$ on our theory (right from the beginning). 
The constraint $p^2 - m^2 = 0$ can be derived from $L_0$ by using the definition of the canonical 
conjugate momenta w.r.t. $x_\mu$ by using the standard formula of the classical mechanics, namely; 
\begin{eqnarray}
p_\mu  = \frac {\partial L_{0}} {\partial \dot x ^\mu}  = \frac {m\,\dot x_{\mu}}{\sqrt {\dot x^{2}}} \quad  \Longrightarrow \quad p_\mu\,p^\mu = m^2, 
\end{eqnarray}
where we have used $\dot x_\mu \, \dot x^\mu = \dot x\cdot \dot x \equiv \dot x^2 \equiv \dot x_0^2 - \dot x_i^2$. 
As we have focused  on the first-order Lagrangian $(L_f)$, it is worthwhile to mention  that,  for {\it this} Lagrangian,  there are {\it two} 
constraints: $\Pi_e \approx 0$ and  $p^2 - m^2 = 0$ where $\Pi_e$ is the canonical conjugate momentum  w.r.t. $e\, (\tau)$. 
The primary constraint ($\Pi_e \approx 0$) and  secondary constraint ($p^2 - m^2 \approx 0$) are the generators for  the 
infinitesimal gauge symmetry transformations [cf. Eq. (17)].

We end this section with the following remarks.
First, the (anti-)BRST invariant CF-type restriction [$B + \bar B + i \,(\bar C \, \dot C - \dot {\bar C} \, C) = 0$] will  be 
computed for all the (non-)interacting, (non-)relativistic and (non-)SUSY systems of 1D particles using the beauty of MBTSA in our Sec. 6. Second, the coupled 
(but equivalent) Lagrangians that respect the (anti-)BRST symmetry transformations [cf. Eq. (22)], are as follows (see e.g. [31] for details):
\begin{eqnarray}
 L_{\bar{b}} & = & p_\mu\,\dot{x}^\mu - \frac{e}{2}\,\big(p^2 - m^2\big) - 
\bar B\,\Big[e\,\dot e - i\,(2\, {\bar C}\,\dot C +  \dot{\bar C}\,C)\Big]+\frac{{\bar B}^2}{2} \nonumber\\
&&- i\,e\,\dot e\,{\bar C}\,\dot C - \,i\,e^2\,\dot{\bar C}\,\dot C  - \dot{\bar C}\,{\bar C}\,\dot C\,C,\nonumber\\
L_{{b}} & = & p_\mu\,\dot{x}^\mu - \frac{e}{2}\,\big(p^2 - m^2\big) + B\,\Big[e\,\dot e -i\, (2\, \dot{\bar C}\, C  
+ {\bar C}\,\dot C)\Big]+ \frac{B^2}{2} \nonumber\\
&&- i\,e\,\dot e\,\dot{\bar C}\,C - \,i\,e^2\,\dot{\bar C}\,\dot C  - \dot{\bar C}\,{\bar C}\,\dot C\,C. 
\end{eqnarray}
At this juncture, it is straightforward to note that, we have the following transformations for $L_{\bar b}$ and $L_{ b}$, 
respectively, under the (anti-)BRST symmetries [cf. Eq. (22)], namely;
\begin{eqnarray}
s_{ab}\,L_{\bar b} & = & \frac{d}{d\,\tau}\,\Big[\bar C\,L_f + {\bar B}^2\,\bar C  
- i\,\bar B\,\dot{\bar C}\,\bar C\,C - e\,\dot e\,\bar B\,\bar C - e^2\,\bar B\,\dot{\bar C}  \Big], \nonumber\\
s_b\,L_b & = &  \frac{d}{d\,\tau}\,\Big[C\,L_f + B^2\,C - i\,B\,\bar C\,\dot C\,C + e\,\dot e\,B\,C  + e^2\,B\,\dot C \Big].
\end{eqnarray}
Thus, the action integrals $S_1 = \int_{-\infty }^{+\infty } d\tau \,L_{\bar b}$ and $S_2 = \int_{-\infty }^{+\infty } d\tau \,L_{b}$
remain invariant under the (anti-)BRST symmetry transformations, respectively, for the physical variables that vanish off as 
$\tau \longrightarrow \pm\, \infty$. Finally, it can be seen that  $L_{\bar b}$   and $L_b$ {\it also} respect the BRST 
and anti-BRST symmetry transformations, respectively, provided we use the validity of the  
(anti-)BRST invariant CF-type restriction: $B + \bar B + i \,(\bar C \, \dot C - \dot {\bar C} \, C) = 0$ that exists on our theory 
(see, e.g. [31] for details). In fact, it turns out that when we apply $s_b$ on $L_{\bar b}$ and $s_{ab}$ on $L_b$,
these Lagrangians transform to the total derivatives {\it plus} terms that vanish when  we invoke the validity of  the 
CF-type restriction. Thus, we note that $L_{ b}$ and $L_{\bar b}$ are {\it equivalent} from the point of view of the symmetry considerations.


\section {(Anti-)BRST symmetries for a Massive Free Spinning Relativistic Particle: (Super)gauge and 1D Diffeomorphism Invariant Model}

The interesting SUSY system of a 1D diffeomorphism (i.e. reparameterization) 
invariant model of a free spinning relativistic particle has been studied
by many people\footnote{Being a prototype example of the SUSY system, the free massive spinning relativistic particle has been studied from 
the point of view of the closed and open gauge algebras where the Batalin-Vilkovsky Lagrangian formulation has been adopted to define the 
(anti-)BRST symmetries and the corresponding conserved charges [43]. This model, with ${\mathcal N}$-extended SUSY, has been quantized 
using the elegant combination of the Batalin-Fradkin-Vilkovsky and the BRST formalism [44]. Using the rigged Hilbert space 
technique [45] and extended phase space, the BRST quantization of {\it this} system has been performed, too [46]. 
However, in our present endeavor, we have used the straightforward and simple Lagrangian of [40] and derived the
proper (anti-)BRST symmetry transformations along with the CF-type restriction (see, e.g. [48] for details).  
The {\it latter} has been not derived, to the best of our knowledge, {\it elsewhere} in the literature.} 
(see, e.g. [40-47] and references therein) from quite different theoretical angels.
In this section, we focus on the nilpotent and absolutely anticommuting (anti-)BRST symmetry transformations, coupled (but equivalent) 
Lagrangians and CF-type restriction corresponding to the  classical, infinitesimal and continuous $(i)$ the 
(super)gauge symmetry transformations, and (ii) the 1D diffeomorphism symmetry transformations. This SUSY system is 
nothing but the model of a free massive  spinning relativistic particle  whose  {\it two} equivalent Lagrangians 
($L_0, \;L_f $)are as follows 
(see, e.g. [32, 40] for details) 
\begin{eqnarray}
&&L_0 = m\,\sqrt {(\dot x^\mu\, + i\,\chi \,\psi^\mu)(\dot x_\mu\,+i\,\chi\, \psi_\mu)} 
+\frac{i}{2}(\psi_\mu \,\dot\psi^\mu - \psi_5 \,\dot \psi_5)  - i\,m\,\chi\,\psi_5,\nonumber\\ 
&&L_f = p_\mu\,\dot x^\mu +\frac{i}{2}(\psi_\mu \,\dot\psi^\mu - \psi_5 \,\dot \psi_5) - \frac {e}{2}\;(p^2 - m^2)
+ i\,\chi \,(p_\mu \,\psi^\mu - m\,\psi_5), 
\end{eqnarray}
where the vectors $x_\mu$ and $\psi_\mu$ are the super partners of each-other [with the fermionic (i.e. $\psi_\mu ^2 = 0, \; \psi_\mu\,\psi_\nu 
+ \psi_\nu\, \psi_\mu = 0$) vector $(\psi_\mu)$ variable]. Both the vector variables are  defined on the D-dimensional 
Minkowskian target  spacetime manifold  {\it but} $x_\mu (\tau)$ and $\psi_\mu (\tau)$
are {\it scalars} w.r.t. the 1D evolution parameter $\tau$ that (i) parameterizes the 1D trajectory 
of the propagation of the free massive spinning particle, and (ii) is responsible for the definition of the  generalized velocities 
$\dot x_\mu  = (dx_\mu/d\tau)$ and $\dot \psi_\mu = (d\psi_\mu/d\tau)$, the pair  $(x_\mu, p^\mu)$ denotes the canonically 
conjugate coordinates and momenta in the D-dimensional flat Minkowskian  target space, fermionic ($\psi_5 ^2 = 0$) variable $\psi_5$ has been invoked 
for the purpose of satisfying the  mass-shell condition $p^2 - m^2 = 0$ where $m$ is the rest mass of the spinning particle, 
$e(\tau)$ is the einbein variable and $\chi  (\tau)$ is its super partner which is fermionic $(\chi^2 = 0)$ in nature. 
All the bosonic variables $(x_\mu, \, p_\mu, \, e)$ commute among themselves as well as with the fermionic 
variables $(\chi, \psi_5, \psi_\mu)$ {\it but} the fermionic variables anticommute (i.e. $\chi\,\psi_\mu + \psi_\mu\, \chi = 0, \; 
\chi\,\psi_5 + \psi_5\,\chi  = 0,\; \psi_5\, \psi_\mu  + \psi_\mu\, \psi_5 = 0$, etc.) among themeselves as well
 as with the fermionic (anti-)BRST symmetry operators [$s_{(a)b}$] which are listed below [cf. Eqs. (39), (45)]. In other words, we {\it also} have
 $\{s_{(a)b}, \;\psi_\mu\} = \{s_{(a)b}, \;\chi\} = \{s_{(a)b}, \;\psi_5\} = 0 $.

It can be checked that the starting Lagrangian for the free massive spinning relativistic particle $(L_0)$ is {\it singular} because we observe that:  
\begin{eqnarray}
W_{\alpha\beta} = \frac  {\partial^2 \, L_0}{\partial \,\dot x^\alpha\,\partial \,\dot x^\beta} 
= \frac {m}{\{(\dot x^\mu\, + i\,\chi \,\psi^\mu)(\dot x_\mu\,+i\,\chi\, \psi_\mu)\}^{{3}/{2}}}\;
\Big[\eta_{\alpha\beta}\;(\dot x_\mu\,+i\,\chi\, \psi_\mu) (\dot x^\mu\, + i\,\chi \,\psi^\mu)\nonumber\\
 -  \,(\dot x_\alpha\,
 + i\,\chi \,\psi_\alpha)(\dot x_\beta \, + i\,\chi\, \psi_\beta) \Big].
\end{eqnarray}
At this stage, it can be noted that the combination $\dot x_\mu\,+i\,\chi\, \psi_\mu$ appears in {\it all} the terms of $W_{\alpha\beta}$. Hence, we choose the  
abbreviation: $\dot X_\mu = \dot x_\mu\,+i\,\chi\, \psi_\mu$. Moreover, we also use the notation: 
$A = \dot X_\mu\, \dot X^\mu \equiv \dot X_0^2\, - \dot X_i ^2$ for 
the Lorentz scalar that is constructed from $\dot X_\mu$. With these notations as the inputs, 
we have the following  concise expression for (32), namely;
\begin{eqnarray}
W_{\alpha\beta} = \frac {m} {(\dot X_\mu \dot X^\mu)^{{3}/{2}}}\;
\Big[\eta_{\alpha\beta}\;\dot X_\mu \dot X^\mu -  \,\dot X_\alpha \, \dot X_\beta\Big]
\equiv  \frac {m} {A^{{3}/{2}}}\;
\Big[\eta_{\alpha\beta}\;A -  \,X_\alpha \, X_\beta\Big].
\end{eqnarray}
It can be explicitly checked that the determinant of $W_{\alpha\beta}$ turns out
 to be {\it zero}. To corroborate this assertion, we note the following for the 2D 
[i.e. (1 + 1)-dimensional] Minkowskian flat target space (where $\alpha, \beta = 0, 1$ for the 
$2\times 2$ matrix form of $W_{\alpha\beta}$), namely; 
\begin{eqnarray}
W_{\alpha\beta} \; = \;
\renewcommand\arraystretch{2}\large{
\begin{pmatrix}
W_{00} \;  & \; W_{01} 
 \\ W_{10}  \;  & \;W_{11} 
\end{pmatrix}
 = 
\renewcommand\arraystretch{2}
\begin{pmatrix}
\frac{{\partial^2 \,L_0}}{{\partial\, \dot  x^0 \, \partial\, \dot x^0}} \;  & \; \frac{{\partial^2 \,L_0}}{{\partial\, \dot  x^0 \, \partial\,\dot  x^1}}
 \\ \frac{{\partial^2 \,L_0}}{{\partial\, \dot  x^1 \, \partial\, \dot x^0}} \;  & \; \frac{{\partial^2 \,L_0}}{{\partial\, \dot  x^1 \, \partial\,\dot  x^1}}
\end{pmatrix}.}
\end{eqnarray}
The explicit form of the above matrix is as follows: 
\begin{eqnarray}
W_{\alpha\beta} \; = \;
\renewcommand\arraystretch{2}\large{
\frac {m}{A^{{3}/{2}}}\;
\renewcommand\arraystretch{2}
\begin{pmatrix}
 (A - \dot X_0^2)\;  & \; -\, \dot X_0\, \dot X_1
 \\ -\, \dot X_0\, \dot X_1 \;  & \; -\, (A + \dot X_1^2)
\end{pmatrix}.}
\end{eqnarray} 
We point out   that the bosonic nature of $\dot X_0$ and $\dot X_1$ allows us to take into account: $[\dot X_0, \, \dot X_1] = 0$.
Now it is straightforward to check  that the det $W_{\alpha\beta} = 0$ (see, e.g. [24, 23] for details) because we have the following explicit
expression for the {\it same}, namely; 
\begin{eqnarray}
\mbox {det} \; W_{\alpha\beta}  = -\, \frac {m^2} {A^{3}}\,A\,  \Big [A + \dot X_1^2 - \dot X_0^2\Big]
 =  -\, \frac {m^2} {A^{2}}\,  \Big [\dot X_0^2 - \dot X_1^2 + \dot X_1^2 - \dot X_0^2\Big] = 0.
\end{eqnarray} 
In our Appendix B, we demonstrate that, for the three (2 + 1)-dimensional (3D) target space, too, the determinant 
of $W_{\alpha\beta}$ (with $\alpha, \beta = 0, 1, 2$) turns out to be zero. 
It is worthwhile to point out that this kind of explicit verification of the {\it singular} 
nature of $L_0$ has {\it not} been done elsewhere. 
 As a consequence of our observation in (36), there are constraints on the 
theory which we elaborate in the next paragraph of our present endeavor taking into consideration  the first-order Lagrangian $L_f$ 
[cf. Eq. (31)].

A close look at $L_f$ [cf. Eq. (31)] demonstrates that there are {\it four} constraints on the free massive 
spinning relativistic  particle of  our theory, namely;
\begin{eqnarray}
\Pi_e \approx 0, \qquad p^2 -m^2 \approx  0, \qquad  \Pi_\chi \approx 0,  \qquad  p^\mu\, \psi_\mu - m\, \psi_5 \approx 0,
\end{eqnarray}
where $\Pi_e $ and $\Pi_\chi$ are the canonical conjugate momenta w.r.t. $e(\tau)$ and $\chi (\tau)$, respectively. 
These (i.e. $\Pi_e \approx 0, \, \Pi_\chi \approx 0$) are the  primary constraints. 
Corresponding to the above primary constraints (i.e. $\Pi_e \approx 0, \, \Pi_\chi \approx 0$), the secondary constraints are:
 $p^2 - m^2 \approx 0$ and $p_\mu\, \psi^\mu - m\, \psi_5 \approx 0$, respectively,
and there are {\it no} further constraints on our theory. It is very obvious that all these 
constraints are first-class (see, e.g. [32] for details) in the terminology of Dirac's prescription for the classification scheme of constraints [23, 24].
The bosonic type constraints ($\Pi_e \approx 0, \, p^2 - m^2 \approx 0$) and fermionic type constraints 
($\Pi_\chi \approx 0, \; p_\mu\, \psi^\mu - m\, \psi_5 \approx 0$) {\it together} lead to the infinitesimal, 
continuous and {\it classical} (super)gauge symmetry transformations ($\delta$) for our theory of 
spinning particle as follows
\begin{eqnarray}
&&\delta\,x_\mu = \xi \, p_\mu +\kappa\, \psi_\mu, \qquad \delta p_\mu = 0,\qquad\delta\psi_\mu = i\,\kappa\,p_\mu,\nonumber\\
&&\delta e = \dot\xi +2\,\kappa\,\chi,\qquad\delta\chi  = i\,\dot\kappa,\qquad \delta\psi_5 = i\,\kappa\,m,
\end{eqnarray}
where $\xi (\tau) $ is the bosonic local gauge symmetry transformation parameter and $\kappa (\tau)$ is the fermionic 
($\kappa^2 = 0 $) supergauge transformation parameter. The {\it latter} is anticommuting 
($\kappa \, \psi_\mu + \psi_\mu\, \kappa = 0, \;  \kappa \, \chi + \chi \, \kappa = 0, \; \kappa \, \psi_5+ \psi_5 \, \kappa = 0$)
in nature with {\it all} the fermionic variables ($\chi, \, \psi_\mu, \, \psi_5$) of our theory and it 
commutes with all the bosonic variables $(e, \, x_\mu, \; p_\mu)$. 
It can be checked that even if $\kappa = 0$, we have a gauge 
symmetry invariance  in the theory that is generated by the bosonic  first-class constraints: $\Pi_e \approx 0, \; p^2 - m^2 \approx 0$. It turns out that 
all the fermionic variables $(\psi_\mu,\, \psi_5,\, \chi)$ do not transform {\it at all} under this gauge symmetry transformations. 
The complete set of off-shell nilpotent $[s_{(a)b}^2 = 0]$ (anti-)BRST symmetry transformations [$s_{(a)b}$] have been quoted in
 our earlier works (see, e.g. [32, 40] for details), corresponding to the classical (super)gauge transformations ($\delta$), as follows:
\begin{eqnarray*}
&& s_{ab}\; x_\mu = {\bar c}\; p_\mu + \bar \beta \;\psi_\mu, \quad\qquad s_{ab}\; e = \dot {\bar c} + 2 \;\bar \beta\; \chi,  
\;\quad\qquad s_{ab} \;\psi_\mu = i \;\bar \beta\; p_\mu,\nonumber\\
&& s_{ab}\; \bar c = - i \;{\bar \beta}^2, \quad s_{ab}\; c = i\; \bar b, \quad s_{ab}\; \bar \beta = 0, 
\;\quad s_{ab} \; \beta = - i\; \gamma, \quad s_{ab}\; p_\mu = 0, \nonumber\\
&& s_{ab} \;\gamma = 0, \quad s_{ab}\; \bar b = 0, \quad s_{ab}\;\chi = i\; \dot {\bar \beta}, 
\quad s_{ab} \; b =  2\; i\; \bar \beta\; \gamma,\quad s_{ab} \,\psi _5 = i\,\bar\beta\,m,
\end{eqnarray*}
\begin{eqnarray}
&&s_b\; x_\mu = c\;p_\mu + \beta \;\psi_\mu, \quad\qquad s_b\; e = \dot c + 2\;\beta\; \chi,  
\quad\qquad s_b\; \psi_\mu = i\;\beta\; p_\mu,\nonumber\\
&& s_b\;c = - i\; \beta^2, \;\quad s_b \;{\bar c} = i\; b, \;\quad s_b \;\beta = 0, 
\;\quad s_b \;\bar \beta = i \;\gamma, \;\quad s_b\; p_\mu = 0,\nonumber\\
&& s_b \;\gamma = 0, \quad s_b \;b = 0, \quad s_b \;\chi = i\; \dot \beta, 
\qquad s_b\; \bar b = - 2\; i\; \beta\; \gamma,\quad s_{b}\, \psi _5 = i\,\beta \, m.
\end{eqnarray}
We point out that, as far as the absolute anticommutativity   property ($s_b\, s_{ab} + s_{ab} \, s_b = 0$)
of the (anti-)BRST symmetry transformations [$s_{(a)b}$] is concerned, we observe that the following explicit computations are true, namely;
\begin{eqnarray}
\{s_b, s_{ab}\}\,x_\mu = i\, (b + \bar b + 2\,\beta\bar\beta)\,p_\mu,\qquad
\{s_b, s_{ab}\}\,e  = i\, \frac{d}{d\tau}\, \big [b + \bar b + 2\,\beta\bar\beta \big ],
\end{eqnarray}
which demonstrate  that the absolute anticommutativity property (i.e. $\{ s_b, \, s_{ab}\} = 0$) is {\it true}
in the above  provided we take into consideration: $b + \bar b + 2\,\beta\bar\beta  = 0$.
The {\it latter} is nothing but the (anti-)BRST invariant  CF-type restriction on our theory.

A few comments are in order. First, the (anti-)ghost variables $(\bar c)c$ are fermionic $(c^2 = \bar c^2 = 0, \; c\, \bar c + \bar c\, c = 0)$
in nature and they are the generalizations of the classical bosonic gauge symmetry transformations parameter $\xi (\tau)$. 
On the other hand, the (anti-)ghost variables $(\bar\beta)\beta$ are bosonic ($\beta^2 \neq 0\, \bar\beta \neq 0$) in nature and they 
are the generalizations of the classical fermionic  $(\kappa^2 = 0)$ SUSY transformation parameter $\kappa (\tau)$. Second, 
the pair $(b, \, \bar b)$ is a set of Nakanishi-Lautrup type auxiliary variables that are present along with $(\bar\beta)\beta$ 
in the CF-type restriction: $b + \bar b + 2\, \bar\beta\, \beta = 0$. Third, the CF-type restriction: $b + \bar b + 2\, \bar\beta\, \beta = 0$
is a {\it physical} restriction because it is (anti-)BRST invariant: $s_{(a)b}\, [b + \bar b + 2\, \bar\beta\, \beta = 0] = 0$ as 
 can be explicitly checked using the (anti-)BRST  transformations quoted in Eq. (39). 
Fourth, it can be checked that for the generic variable $\phi (\tau)$, we have: $\{s_b, \; s_{ab}\} \, \phi = 0$
where $\phi  = p_\mu, \, b,\, \bar b,\,  \bar c, \, c,\, \gamma, \, \beta, \, \bar\beta, \psi_\mu,\, \psi_5$
which shows that the absolute commutativity  for the generic variable $\phi\,(\tau)$ is satisfied automatically. 
Fifth, the CF-type restriction: $b + \bar b + 2\, \bar\beta\, \beta = 0$ has been derived by the supersymmetrization of the
horizontality condition in our earlier work [48]. Sixth, the (anti-)BRST 
 transformations (39) are the symmetry  transformations  for the action integrals corresponding to the coupled 
(but equivalent) Lagrangians [32, 48]
\begin{eqnarray}
L_{\bar b} &=& L_f + {\bar b}^2 - \bar b\,({\dot e} - 2\, {\bar \beta}\,\beta) - i\,\dot{\bar c} \, {\dot c} + {\bar \beta}^2 \, {\beta}^2 
+ 2\, i\, \chi\,(\beta\,\dot{\bar c} - \bar{\beta} \, \dot c)  + 2\,e\,(\dot{\bar \beta} \, \beta - \gamma\,\chi) \nonumber\\
&+& 2\,\gamma\,(\beta \, \bar c - \bar\beta \, c) + m \, (\bar\beta \, \dot\beta - \dot{\bar \beta}\, \beta + \gamma \, \chi)
- \dot \gamma \, \psi_5,
\end{eqnarray}
\begin{eqnarray}
 L_b & = & L_f + b^2 +b\,({\dot e} + 2\, {\bar \beta}\,\beta) - i\,\dot{\bar c} \, {\dot c} + {\bar \beta}^2 \, {\beta}^2 
+ 2\, i\, \chi\,(\beta\,\dot{\bar c} - \bar{\beta} \, \dot c)  - 2\,e\,(\bar\beta \, \dot\beta + \gamma\,\chi) \nonumber\\
&+& 2\,\gamma\,(\beta \, \bar c - \bar\beta \, c) + m \, (\bar\beta \, \dot\beta - \dot{\bar \beta}\, \beta + \gamma \, \chi) 
- \dot \gamma \, \psi_5, 
\end{eqnarray}
where $L_f$ is the first-order Lagrangian (31) and $\gamma$ is an auxiliary fermionic $(\gamma^2 = 0)$ variable.
To be more precise, it can be explicitly  checked  that $L_{\bar b}$ and $L_b$ transform under the off-shell nilpotent, 
continuous and infinitesimal {\it quantum} (anti-)BRST symmetry transformations as the total derivatives w.r.t. the evolution parameter $\tau$, namely; 
\begin{eqnarray}
s_{ab} \, L_{\bar b} = \frac{d}{d\,\tau}\,\Big[\frac{\bar c}{2}\,(p^2 + m^2) + \frac{\bar\beta}{2}\,(p_\mu \, \psi^\mu + m \, \psi_5) 
- \bar b \, (\dot{\bar c} + 2\,\bar\beta \, \chi)\Big],
\end{eqnarray}
\begin{eqnarray}
s_b \, L_b = \frac{d}{d\,\tau}\,\Big[\frac{c}{2}\,(p^2 + m^2) + \frac{\beta}{2}\,(p_\mu \, \psi^\mu + m \, \psi_5)
+ b \, (\dot c + 2\,\beta\, \chi)\Big].   
\end{eqnarray}
Hence, the above anti-BRST and BRST symmetries are the {\it perfect} symmetries of the action integrals 
corresponding to $L_{\bar b}$ and $L_{b}$, respectively. We would like to lay emphasis on the fact that {\it both} the Lagrangians 
$L_{\bar b}$ and $L_{b}$  respect BRST and anti-BRST symmetry transformations, too, provided we invoke the sanctity 
and validity of the (anti-)BRST invariant CF-type restriction: $b + \bar b + 2\, \bar\beta\, \beta = 0$ on our theory
(see, e.g. [32] for details).

We dwell a bit now on the {\it classical} reparameterization (i.e. 1D diffeomorphism) symmetry transformations [cf. Eq. (46) below]
of the  variables of $L_f$ and their generalizations to the {\it quantum} level within the framework of BRST formalism. The 
complete set of off-shell nilpotent [$s_{(a)b}^2 = 0$] (anti-)BRST symmetry transformations (corresponding to the classical 1D diffeomorphism
symmetry transformations) are as follows (see, e.g. [32] for details)
\begin{eqnarray*} 
&& s_{ab} \;\psi_\mu = {\bar C}\; {\dot\psi}_\mu,  \qquad s_{ab}\; p_\mu = {\bar C}\; {\dot p}_\mu, \qquad
s_{ab}\; e = \frac{d}{d\,\tau}\,({\bar C}\,e), \quad s_{ab}\; x_\mu = {\bar C}\; {\dot x}_\mu, \nonumber\\  
&&s_{ab}\; \bar C = \bar C \, \dot{\bar C}, \qquad 
s_{ab}\;\chi = \frac{d}{d\,\tau}\,({\bar C}\,\chi),\quad s_{ab} \,\psi _5 = {\bar C}\; {\dot\psi}_5,  \qquad 
s_{ab}\; C = i\,\bar B, \nonumber\\
&& s_{ab}\; \bar B = 0, \qquad\qquad s_{ab} \; B =  {\dot B}\,{\bar C} - B\,\dot{\bar C},
\end{eqnarray*}
\begin{eqnarray}
&& s_{b}\; x_\mu = {C}\; {\dot x}_\mu, \qquad s_{b}\; p_\mu = {C}\; {\dot p}_\mu, \qquad
s_{b}\; e = \frac{d}{d\,\tau}\,({C}\,e), \qquad s_{b} \;\psi_\mu = {C}\; {\dot\psi}_\mu, \nonumber\\  
&&s_{b} \,\psi _5 = {C}\; {\dot\psi}_5, \qquad 
s_{b}\;\chi = \frac{d}{d\,\tau}\,({C}\,\chi), \qquad s_{b}\; \bar C = i\,B, \;\;\; \qquad 
 s_{b}\;  B = 0,\nonumber\\
&&s_{b}\; C = C\,{\dot C},  \qquad\qquad s_{b} \; \bar B =  \dot{\bar B}\,{C} - \bar B\,\dot{C},
\end{eqnarray} 
where the {\it classical}, infinitesimal and continuous reparameterization (i.e. 1D diffeomorphism) 
symmetry transformations $(\delta_{r} )$ for the variables of $L_f$ are as follows:
\begin{eqnarray}
&&\delta_{r} \, x_\mu = \varepsilon  \, {\dot x}_\mu, \qquad \qquad \delta_{r} \, p_\mu = 
\varepsilon  \, {\dot p}_\mu, \qquad \qquad \delta_{r} \, \psi_\mu = \varepsilon  \, {\dot \psi}_\mu,\nonumber \\
&&\delta_{r} \,\psi_5 = \varepsilon \, {\dot \psi}_5, \qquad \qquad  \delta_{r} \, e = \frac{d}{d\tau}(\varepsilon \,e), \qquad \qquad 
\delta_{r} \, \chi = \frac{d}{d\tau}(\varepsilon \,\chi).
\end{eqnarray}
Here $\varepsilon (\tau)$ is the infinitesimal parameter in: $\tau \longrightarrow  \tau - \varepsilon (\tau)$.
We would like to point out [40] that the {\it classical}, continuous and infinitesimal (super)gauge symmetry 
transformations $(\delta)$ [cf. Eq. (38)] can be derived from the above reparameterization symmetry transformations 
$(\delta_r)$ if we (i) identify: $\xi(\tau) = e(\tau)\, \varepsilon (\tau), \, \kappa(\tau) = -\, i\, \chi(\tau)\,\varepsilon  (\tau)$, and (ii)
use the equations of motion: $\dot p_\mu = 0, \, \dot x_\mu = e\, p_\mu - i\, \chi\, \psi_\mu, \;  \dot\psi_\mu = \chi\, p_\mu, \, 
\dot \psi_5 = m\, \chi$.
As mentioned  earlier, in the complete set of (anti-)BRST symmetry transformations (45), the fermionic 
$(C^2 = \bar C^2 = 0, \; C\, \bar C + \bar C\, C = 0)$ (anti-)ghost variables $(\bar C) C$ are the generalizations of the
{\it classical}, infinitesimal and continuous reparameterization symmetry transformation parameter $\varepsilon  (\tau)$ and the 
pair $(B, \bar B)$ is a set of Nakanishi-Lautrup auxiliary variables which are responsible for 
the absolute anticommutativity  of the (anti-)BRST symmetry transformations (45) as they appear in the CF-type restriction:
$B + \bar B + i\,(\bar C\,\dot C - \dot{\bar C}\,C) = 0$ along with the fermionic (anti-)ghost variables $(\bar C)C$. For instance, it can be checked that we have
\begin{eqnarray}
&&\{s_b, s_{ab}\}\,S = i\,\big[B + {\bar B} + i\,({\bar C}\,{\dot C} - \dot{\bar C}\,C)\big]\,{\dot S}, \nonumber\\
&&\{s_b, s_{ab}\}\,\Phi = i\,\frac{d}{d\,\tau}\Big[\big\{B + {\bar B} + i\,({\bar C}\,{\dot C} - \dot{\bar C}\,C\big\}\,\Phi\Big], \nonumber\\
&&\{s_b, s_{ab}\}\,\Psi = 0,
\end{eqnarray}
where $ S = (x_\mu ,\,p_\mu ,\,\psi_\mu ,\,\psi_5 )$,  $\Phi= e (\tau),\, \chi (\tau)$ and $ \Psi = B, \bar B, C, \bar C$ are the generic variables.
Furthermore, the CF-type restriction is {\it also} at the heart of the existence of the (anti-)BRST invariant coupled (but equivalent) Lagrangians as quoted below:
\begin{eqnarray}
&&L_b = L_f  + B\,\Big[e\,\dot e -i\, (2\, \dot{\bar C}\, C  + {\bar C}\,\dot C)\Big]+ \frac{B^2}{2} -\,i\,e^2\,\dot{\bar C}\,\dot C 
- i\,e\,\dot e\,\dot{\bar C}\,C - \dot{\bar C}\,{\bar C}\,\dot C\,C, \nonumber\\
&&L_{\bar b} = L_f  - \bar B\,\Big[e\,\dot e - i\,(2\, {\bar C}\,\dot C +  \dot{\bar C}\,C)\Big]
+\frac{{\bar B}^2}{2}  -\,i\,e^2\,\dot{\bar C}\,\dot C - i\,e\,\dot e\,{\bar C}\,\dot C - \dot{\bar C}\,{\bar C}\,\dot C\,C, 
\end{eqnarray}
where $L_f$ is the first-order Lagrangian of (31). It is straightforward to check that $L_{\bar b}$ and $L_b$ transform under the (anti-)BRST
symmetry transformations $[s_{(a)b}]$ to  the total derivatives w.r.t. the evolution parameter $(\tau)$, namely; 
\begin{eqnarray}
&&s_{ab}\,L_{\bar b} = \frac{d}{d\,\tau}\Big[\bar C\,L_f - e^2\,\bar B\,\dot{\bar C} - e\,\dot e\,\bar B\,\bar C 
- i\,\bar B\,\dot{\bar C}\,\bar C\,C + {\bar B}^2\,\bar C\Big], \nonumber\\
&&s_b\,L_b = \frac{d}{d\,\tau}\Big[C\,L_f + e^2\,B\,\dot C + e\,\dot e\,B\,C - i\,B\,\bar C\,\dot C\,C + B^2\,C\Big].
\end{eqnarray}
As a consequence, the action integrals $S_1 = \int_{-\infty }^{+\infty } d\tau \,L_{\bar b}$ and $S_2 = \int_{-\infty }^{+\infty } d\tau \,L_{b}$
are invariant (i.e. $s_{ab} S_1 = 0, \; s_{b} S_2 = 0$) under the anti-BRST  and BRST transformations, respectively, for the 
physical variables that vanish off as $\tau \longrightarrow \pm\, \infty$.
We further  point out that $L_{\bar b}$ and $L_b$ {\it also} respect BRST and anti-BRST symmetry transformations, respectively, 
provided we use the validity of the CF-type restriction: 
$B + \bar B + i\,(\bar C\,\dot C - \dot{\bar C}\,C) = 0$ [23].

We conclude  this section with the following remarks. First of all, we point out  that the (anti-)BRST invariant Lagrangians
 [cf. Eqs (41), (42), (48)] are coupled
(but equivalent) for the generalizations of the (i) classical, infinitesimal and continuous (super)gauge symmetry transformations,
and (ii)  classical, infinitesimal and continuous 1D diffeomorphism symmetry transformations. 
Second, the CF-type restriction: $b + \bar b +2\, \bar \beta\, \beta = 0$ reduces to its {\it trivial} form
(i.e. $b + \bar b = 0$) for the scalar relativistic particle (which is the limiting case of a free massive spinning relativistic  particle)
where there is existence of a {\it single} Lagrangian 
that respects {\it both} (i.e. BRST and  anti-BRST) symmetry transformations (see, e.g. [31] for details). 
Third, the CF-type restriction for the (anti-)BRST symmetries,
corresponding to the classical, infinitesimal and continuous 1D diffeomorphism symmetry transformations, is {\it universal}
and, for the 1D reparameterization invariant systems, it is always:  $B + \bar B + i\,(\bar C\,\dot C - \dot{\bar C}\,C) = 0$. Third, the {\it universal} nature 
of the CF-type restriction can be explained by the MBTSA to BRST formalism [7, 8] (cf. Sec. 6 below, too).


\section {Gauge and 1D Diffeomorphism Invariant Interacting Model of a Scalar Relativistic Particle: 
(Anti-) BRST Symmetries and CF-type Restriction}

This section is divided into {\it three} subsections where we lay emphasis on the {\it classical} as well as {\it quantum} versions of the gauge 
and diffeomorphism symmetries.


\subsection {Classical Gauge and Reparameterization Symmetries}

We begin with {\it two} equivalent Lagrangians ($L_0, \; L_f$)
that describe the motion of a  scalar relativistic particle 
in interaction with an electromagnetic field that is in the background. These Lagrangians are as follows 
(see, e.g. [49, 23, 36] for details)
\begin{eqnarray}
&& L_{0} = m \; (\dot x^2)^{1/2} - {\displaystyle \frac{1}{2}}
F_{\mu\nu} \,x^\mu \dot x^\nu, \nonumber\\
&& L_{f} = p_\mu \,\dot x^\mu - {\displaystyle \frac{1}{2}}\, F_{\mu\nu} \,x^\mu \dot x^\nu -
{\displaystyle \frac{1}{2}}\;e\; (p^2 - m^2), 
\end{eqnarray} 
where $L_0$ is the Lagrangian with the square root, $L_f$ is the first-order Lagrangian 
where the dynamical variables are maximum in number and these are ($x_\mu, \, p_\mu, \, e$).
 Here the rest mass $m$ and background electromagnetic field $F_{\mu\nu}$
 are independent of the evolution parameter $\tau$ that characterizes the trajectory of the particle in the D-dimensional 
 Minkowski spacetime manifold (with $\mu, \, \nu, \, \lambda,... = 0, 1, 2,...,D - 1$). For our further discussions, we shall focus 
on the first-order Lagrangian which has maximum number of dynamical variables and there is {\it no} variable  in the denominator. 
We also have  $\dot x^\mu = (dx^\mu/d\tau)$ as the generalized velocity for the target space coordinates.

The first-order Lagrangian ($L_f$) is endowed with the first-class constraints: $\Pi_e \approx 0$ and $p^2 - m^2 \approx 0$
[where $\Pi_e$ is the canonical conjugate momentum w.r.t. the variable $e(\tau)$ and $p^2 - m^2  \approx  0$ is the mass-shell condition]. 
These first-class constraints generate an infinitesimal, continuous, classical gauge symmetry transformation (see, e.g. [49, 25, 26]) 
\begin{eqnarray}
\delta_g x_\mu = \xi p_\mu, \qquad
\delta_g p_\mu = - \xi F_{\mu\nu} p^\nu, \qquad \delta_g F_{\mu\nu} = 0,
\qquad \delta_g e = \dot \xi,
\end{eqnarray}  
where $\xi(\tau)$ is the infinitesimal gauge symmetry transformation parameter. It is straightforward 
to check that $L_f$ transforms to a  total derivative under the infinitesimal  {\it classical} gauge symmetry transformations ($\delta_g$): 
\begin{eqnarray}
\delta_g L_f = \frac {d}{d\tau}\Big[ \frac {\xi}{2}\, \big(p^2 + m^2 - F_{\mu\nu}\,x^\mu\,p^\nu \big)\Big].
\end{eqnarray}  
Hence, the action integral  $S = \int_{- \, \infty}^{+ \, \infty} d\tau \, L_f$ remains invariant ($\delta_g\, S = 0$) under $\delta_g$. 
The above first-class constraints exist because the starting Lagrangian $L_0$ is {\it singular}. The latter property can be proven along the similar  
lines of arguments,  as we have done for the {\it free} scalar relativistic 
particle in Sec. 3 [cf. Eqs. (26), (27) and Appendix A for details].

In addition to the above {\it classical} gauge symmetry transformations ($\delta_g$) that have been generated by the first-class constraints 
on our theory, we {\it also} have  a continuous, infinitesimal and {\it classical} reparameterization transformation $(\delta_r)$
due to the infinitesimal {\it basic} 1D diffeomorphism (i.e. reparameterization) transformations of the evolution parameter: 
$\tau \longrightarrow \tau - \varepsilon (\tau)$ that characterizes  the trajectory of the particle. These transformations  are 
\begin{eqnarray}
\delta_r x_\mu = \varepsilon \,\dot x_\mu, \qquad \delta_r p_\mu = \varepsilon \,\dot p_\mu, \qquad \delta_r F_{\mu\nu} = 0,
\qquad \delta_r e = {\displaystyle \frac{d}{d\tau}}\; [\varepsilon \,e],\qquad \delta_r L_f = \frac {d}{d\tau}\Big[ \varepsilon\, L_f \Big], 
\end{eqnarray} 
which show that the action integral $S = \int_{- \, \infty}^{+ \, \infty} d\tau \, L_f$ remains invariant ($\delta_r\, S = 0$)
for the physically well-defined infinitesimal parametrization transformation parameter $\varepsilon (\tau)$ and the variables of the first-order
Lagrangian $(L_f)$. It is worth pointing out that the infinitesimal gauge and reparameterization symmetries are {\it equivalent} [49]  if (i) 
we identify the gauge transformation parameter $\xi (\tau)$ with the reparameterization symmetry parameter $\varepsilon (\tau)$ 
as: $\xi (\tau)  = e(\tau)\, \varepsilon (\tau)$, and (ii) use the equations of motion:
$\dot x_\mu = e \,p_\mu,  \;\dot p_\mu  = F_{\mu\nu} \dot x^\nu  \equiv -\, e\, F_{\mu\nu}\,p^\nu.$
We also note that $x_\mu$ and $p_\mu$ transform like the Lorentz scalars under the infinitesimal 1D diffeomorphism (i.e. 
reparameterization) symmetry transformations but $e(\tau)$ transforms as the scalar density (in 1D). A close look at 
the EL-EoM: $\dot p_\mu  = F_{\mu\nu}\,\dot x^\nu$ demonstrates that the motion of the particle is {\it not}  a {\it free} motion (i.e. $\dot p_\mu \ne 0$) 
unless  $F_{\mu\nu} = 0$. Thus, it is an {\it interacting} system of a massive  scalar relativistic particle.

We wrap-up  this subsection with the following remarks. First of all, as  pointed out earlier, all the 1D diffeomorphism invariant 
systems  (discussed in our present endeavor) are gauge as well as reparameterization invariant. Second, the infinitesimal gauge and reparameterization 
symmetries are found to be {\it equivalent} under specific identifications and use of equations of motion. Third, the generator [25, 26]
for the infinitesimal gauge symmetry transformations ($\delta_g$) for our present system (in terms of the first-class constraints) is    
\begin{eqnarray}
G = \dot \xi\, \Pi_e + \frac {1}{2}\, \xi\, (p^2 - m^2),
\end{eqnarray} 
where the first-class constraints $\Pi_e \approx 0$ and $-\, (1/2)\,(p^2 - m^2) \approx 0$ are present.  For the generic variable 
$\Phi$, the above generator leads to the following gauge transformations 
\begin{eqnarray}
\delta_g\, \Phi = -\,i\, [\Phi, G], \qquad \Phi  = e,\,x_\mu,\, p_\mu, 
\end{eqnarray}
where the canonical commutators $[e, \,\Pi_e] = i, \; [x^\mu, \,p_\nu] = i\, \delta_\nu ^\mu$ and $[p_\mu, \,p_\nu] = -\, i\, F_{\mu\nu}$ have to be used 
(see, e.g. [49] for details) for the precise and explicit derivations.\\

\subsection{Quantum (Anti-)BRST Symmetries Corresponding to the Classical Gauge Symmetries: A Brief Sketch}

We elevate the {\it classical}, infinitesimal and continuous gauge symmetry transformations (51) to 
their counterparts off-shell nilpotent, infinitesimal and continuous {\it quantum} (anti-)BRST symmetry 
transformations  $(s_{(a)b})$ as 
\begin{eqnarray}
&&s_{ab} x_\mu = \bar c\; p_\mu,  \qquad\; s_{ab} p_\mu = - \bar c\;
F_{\mu\nu}\; p^\nu, \qquad \; s_{ab} \bar c = 0, \nonumber\\
&& s_{ab}e = \dot {\bar c}, \qquad s_{ab} c = - i\; b, \qquad s_{ab} b = 0,\qquad s_{ab} F_{\mu\nu} = 0,\nonumber\\
&& s_b x_\mu = c\; p_\mu,  \qquad\;\;
s_b p_\mu = - c\; F_{\mu\nu}\; p^\nu, \qquad\;\;s_b c = 0, \nonumber\\
&& s_b e = \dot c, \qquad\; s_b \bar c = i\; b, \;\qquad s_b b = 0, \qquad \; s_{b} F_{\mu\nu} = 0,
\end{eqnarray} 
under which the gauge-fixed Lagrangian $(L_b)$   
 \begin{eqnarray}
L_b = p_\mu \,\dot x^\mu
- {\displaystyle \frac{1}{2}} \,F_{\mu\nu} \,x^\mu \,\dot x^\nu
- {\displaystyle \frac{1}{2}}\;e\; (p^2 - m^2)
+ b \;\dot e  + {\displaystyle \frac{1}{2}}\; b^2 - i \;\dot {\bar c} \;\dot c,
 \end{eqnarray} 
that also  contains the fermionic $(c^2 = \bar c^2 = 0, \, c\, \bar c + \bar c \, c = 0)$ (anti-)ghost variables $(\bar c)c$ 
and the  Nakanishi-Lautrup type auxiliary variable $b$, transforms to the total derivatives:
\begin{eqnarray}
&& s_{ab} \; L_b = {\displaystyle \frac{d}{d\tau}\;
\Bigl [\; \frac{\bar c}{2} \bigl ( \; p^2 + m^2 -  F_{\mu\nu}\;
x^\mu\; p^\nu \bigr ) + b\; \dot {\bar c}} \Bigr ], \nonumber\\
&& s_b \; L_b = {\displaystyle \frac{d}{d\tau}\;
\Bigl [\; \frac{c}{2} \bigl ( \; p^2 + m^2 - F_{\mu\nu}\;
x^\mu\; p^\nu \bigr ) + b\; \dot c} \Bigr ]. 
\end{eqnarray} 
A close look at the (anti-)BRST symmetry transformations on the einbien variable [cf. Eq. (56)]
demonstrates that this Lagrange multiplier variable is like the ``gauge" variable for our {\it interacting} theory.
As a consequence of our observations in (58), the action integral $S = \int_{-\, \infty}^{+\, \infty} d \tau\, L_b$ remains invariant 
$(s_{(a)b}\, S = 0)$ under the off-shell nilpotent (anti-) BRST symmetry transformations 
 for the physically well-defined variables that vanish off as $\tau \longrightarrow \pm\, \infty$.
According to Noether's theorem, the above continuous (anti-)BRST symmetry transformations lead to the derivations  of the 
conserved (anti-)BRST charges $[Q_{(a)b}]$ as 
\begin{eqnarray}
Q_{ab} = {\displaystyle b \,\dot {\bar c} + \frac{1}{2} \,(p^2 - m^2)\, \bar c} \; \equiv \; b\, \dot {\bar c} - \dot b\, \bar c,\qquad
{\displaystyle Q_b = b \,\dot c + \frac{1}{2} \,(p^2- m^2)\, c} \; \equiv \; b\,\dot c - \dot b\, c,
\end{eqnarray}
where, in the derivation of the final form of $Q_{(a)b}$, we have used the EL-EoM w.r.t. $e(\tau)$.  
The above conserved charges are the generators for the (anti-)BRST symmetry transformations (56) which are responsible 
for their derivation. It is straightforward to check that, for a generic variable $\Psi = x_\mu, \, p_\mu, \, 
e, \, c, \, \bar c, \, b$, we have $s_{(a)b}\, \Psi = -\,i\,  [\Psi, \, Q_{(a)b}]_{\pm}$ where the subscripts $(+)-$, on the 
square bracket, stand for the (anti)commutator for the given generic variable $\Psi$ of our theory being (fermionic)bosonic in nature.

We conclude  this subsection with the following remarks. First of all, the (anti-)BRST symmetry transformations 
are {\it not} only off-shell nilpotent $(s_{(a)b}^2 = 0)$ of order two {\it but} they are also absolutely  anticommuting
$(s_b\, s_{ab} + s_{ab}\, s_b = 0)$ in their operator form. Second, the conserved (anti-)BRST charges $[Q_{(a)b}]$, that are 
derived from the Noether theorem by using the off-shell nilpotent and absolutely anticommuting, infinitesimal and continuous 
(anti-)BRST symmetry transformations, are {\it also} off-shell nilpotent $[Q^2_{(a)b} = 0]$ of order two and 
absolutely anticommuting in nature which can be easily verified by using the following straightforward theoretical  technique, namely; 
\begin{eqnarray}
s_b\, Q_b = -\, i\, \{ Q_b, \, Q_b \} = 0, \qquad s_{ab}\, Q_{ab} = -\, i\, \{ Q_{ab}, \, Q_{ab} \} = 0, \nonumber\\
s_{ab}\, Q_b = -\, i\, \{ Q_b, \, Q_{ab} \} = 0, \qquad s_{b}\, Q_{ab} = -\, i\, \{ Q_{ab}, \, Q_{b} \} = 0, 
\end{eqnarray} 
where $(i)$  the conserved (anti-)BRST charges are treated as the generators for the continuous
 (anti-)BRST symmetry transformations $[s_{(a)b}]$, and (ii) the entries of the  l.h.s. of equation (60)
are explicitly computed using the expressions for the (anti-)BRST charges [cf. Eq. (59)]. 
Third, for the Lagrangian $L_b$, it is clear that BRST-quantization can be performed because {\it all} the variables of the 
theory have their corresponding  canonical conjugate momenta. 
Fourth, the Nakanishi-Lautrup type auxiliary variable $b$ has been invoked to
linearize the square gauge-fixing term (i.e. $b \;\dot e  + \frac{1}{2}\; b^2 = -\, \frac{1}{2}\, \dot e^2$).
Finally, due to the presence of the (anti-)ghost variables, it is clear that the {\it unitarity} 
of our (anti-)BRST invariant theory is guaranteed at the {\it quantum}  level.


\subsection{Quantum (Anti-)BRST Symmetries Corresponding to the Classical  Reparameterization Symmetries: A Synopsis}

In this subsection, we promote the {\it classical}, continuous and infinitesimal 1D diffeomorphism (i.e. reparameterization) symmetry transformations 
$(\delta_r)$ to their counterparts {\it quantum} infinitesimal and continuous off-shell nilpotent (anti-)BRST transformations $[s_{(a)b}]$
\begin{eqnarray}
&&s_{ab}\,x_\mu = \bar C\,\dot x_\mu,\quad s_{ab}\,p_\mu = \bar C\,\dot{p_\mu},\quad s_{ab}\,e =
 \frac{d}{d\,\tau}\,(\bar C\, e),\quad s_{ab} F_{\mu\nu} = 0,\nonumber\\
&&s_{ab}\,C = i\, \bar B,\quad s_{ab}\, \bar C = \bar C\, \dot {\bar C}, \quad s_{ab}\,\bar B = 0, \quad  s_{ab}\, B =
 \dot {B}\, \bar C - B\, \dot{\bar C},\nonumber\\
&&s_{b}\,x_\mu = C\,\dot x_\mu,\quad s_{b}\,p_\mu = C\,\dot{p_\mu},\,\quad s_{b}\,e = \frac{d}{d\,\tau}\,( C\, e), \quad s_{b}F_{\mu\nu} = 0,\nonumber\\
&&s_{b}\,\bar C = i\, B,\quad s_{b}\, C = C\, \dot C, \quad s_{b}\, B = 0, \quad  s_{b}\, \bar B = \dot {\bar B}\, C - \bar B\, \dot C,
\end{eqnarray}
where the fermionic ($\bar C^2 = C^2 = 0, \; C\, \bar C + \bar C\, C = 0$) variables are the (anti-)ghost $(\bar C)C$ variables 
which are the generalizations of the infinitesimal diffeomorphism transformation parameter $\varepsilon (\tau)$ that are present in the 
transformations ($\delta_r$). In addition, we have the Nakanishi-Lautrup auxiliary variables $(\bar B)B$. It is straightforward 
to note that the above (anti-)BRST symmetry transformations [$s_{(a)b}$] are off-shell nilpotent $(s_{(a)b}^2 = 0)$ of order two. 
As far as the absolute anticommutativity property is concerned, we observe the following: 
\begin{eqnarray}
&&\{s_b, \, s_{ab}\}\, \phi_\mu = i\, [B + \bar B + i\, (\bar C\, \dot C - \dot {\bar C}\, C)]\, \dot \phi_\mu, \qquad  \phi_\mu = x_\mu, \, p_\mu, \nonumber\\
&&\{s_b, \, s_{ab}\}\, e = i\, \frac {d}{d\tau}\Big[\big\{B + \bar B + i\, (\bar C\, \dot C - \dot {\bar C}\, C)\big\}\, e\Big], \nonumber\\
&&\{s_b, \, s_{ab}\}\, \Phi = 0, \qquad \Phi = B,\, \bar B,\, C,\, \bar C.
\end{eqnarray}
Thus, it is evident that the property of absolute anticommutativity is satisfied   if and only if we invoke the 
following Curci-Ferrari (CF) type restriction: 
\begin{eqnarray}
B + \bar B + i\, (\bar C\, \dot C - \dot {\bar C}\, C) = 0.
\end{eqnarray}
It turns out that this CF-type restriction is an (anti-)BRST invariant 
(i.e. $s_{(a)b} \, [B + \bar B + i\, (\bar C\, \dot C - \dot {\bar C}\, C)] = 0$) quantity
provided the validity of the CF-type restriction is taken into account. Hence, it is a {\it physical} restriction on our theory.
It also respects a discrete set of symmetry transformations as pointed out earlier [cf. Eq. (25)].

The (anti-)BRST transformations (61) are the symmetry transformations for the following coupled (but equivalent) Lagrangians 
\begin{eqnarray}
 L_{\bar{b}} & = & p_\mu\,\dot{x}^\mu - {\displaystyle \frac{1}{2}} \,F_{\mu\nu} \,x^\mu \,\dot x^\nu - \frac{e}{2}\,\big(p^2 - m^2\big) - 
\bar B\,\Big[e\,\dot e - i\,(2\, {\bar C}\,\dot C +  \dot{\bar C}\,C)\Big] \nonumber\\
&& + \frac{{\bar B}^2}{2} - i\,e\,\dot e\,{\bar C}\,\dot C - \,i\,e^2\,\dot{\bar C}\,\dot C  - \dot{\bar C}\,{\bar C}\,\dot C\,C,\nonumber\\
L_{{b}} & = & p_\mu\,\dot{x}^\mu- {\displaystyle \frac{1}{2}} \,F_{\mu\nu} \,x^\mu \,\dot x^\nu 
- \frac{e}{2}\,\big(p^2 - m^2\big) + B\,\Big[e\,\dot e -i\, (2\, \dot{\bar C}\, C  
+ {\bar C}\,\dot C)\Big] \nonumber\\
&& + \frac{B^2}{2} - i\,e\,\dot e\,\dot{\bar C}\,C - \,i\,e^2\,\dot{\bar C}\,\dot C  - \dot{\bar C}\,{\bar C}\,\dot C\,C, 
\end{eqnarray}
because we observe the following symmetry transformations for the above Lagrangians under the (anti-)BRST symmetry transformations 
\begin{eqnarray}
s_{ab}\,L_{\bar b} & = & \frac{d}{d\,\tau}\,\Big[\bar C\,L_f + {\bar B}^2\,\bar C  
- i\,\bar B\,\dot{\bar C}\,\bar C\,C - e\,\dot e\,\bar B\,\bar C - e^2\,\bar B\,\dot{\bar C}  \Big], \nonumber\\
s_b\,L_b & = &  \frac{d}{d\,\tau}\,\Big[C\,L_f + B^2\,C - i\,B\,\bar C\,\dot C\,C + e\,\dot e\,B\,C  + e^2\,B\,\dot C \Big],
\end{eqnarray}
where $L_f$ is the first-order Lagrangian of the equation (50). Hence, the action integrals 
$S_1 = \int_{-\, \infty}^{+\, \infty} d \tau\, L_{\bar b}$ and $S_2 = \int_{-\, \infty}^{+\, \infty} d \tau\, L_b$
remain invariant ($s_{ab} S_1 = 0,\; s_b S_2 = 0$) for the physical variables that vanish off as $\tau \longrightarrow \pm\,\infty$.
Both the above Lagrangians are coupled and equivalent because of the presence of the (anti-)BRST invariant CF-type restriction on our theory.
For instance, it can be checked that if we apply $s_b$ on $L_{\bar b}$ and $s_{ab}$ on $L_b$, we find that the Lagrangians 
transform to the total derivatives {\it plus} terms that vanish due to the validity of the CF-type restriction:
$B + \bar B + i\, (\bar C\, \dot C - \dot {\bar C}\, C) = 0$. Thus, we conclude that, {\it if} the whole theory is considered
 on a submanifold of the total quantum Hilbert space of variables where the CF-type restriction is valid, we find that
 (i) both the coupled Lagrangians  $L_{\bar b}$ and  $L_{b}$ respect {\it both} (i.e. BRST and anti-BRST)
 symmetry transformations, and (ii) the (anti-)BRST symmetry transformations satisfy the property of absolute 
 anticommutativity.

We wrap-up this subsection with the remark that, as far as the diffeomorphism symmetry is concerned, 
our discussions on the {\it free} scalar relativistic particle in Sec. 3 are {\it exactly} same as the {\it ones}
that are valid for the {\it interacting} scalar relativistic particle where the electromagnetic field remains 
in the background.  In fact, a close look at the equations (29) and (30) in the context of a {\it free} scalar 
relativistic particle and our present equations (64) and (65)  establish that they are {\it exactly} 
the same {\it except} that  the key difference lies in the fact that 
the first-order Lagrangians of the above {\it two} systems are different. The difference is only due to the 
presence of the term $[-\,(1/2)\,F_{\mu\nu}\,x^\mu\,\dot x^\nu]$ in the first-order Lagrangian of the {\it interacting}  scalar 
relativistic particle. A detailed discussion on the supervariable approach to the 1D diffeomorphism 
invariant model of a {\it free} scalar particle has been performed in our earlier work [31]. 
These discussions are valid for our present  {\it interacting} model, too, as far as the 1D diffeomorphism symmetry transformation is concerned.\\


\section{CF-Type Restriction for the Reparameterization Invariant 1D Models: MBTSA to BRST Formalism}

In this section, we derive the CF-type restriction: $B + \bar B + i\, (\bar C\, \dot C - \dot {\bar C}\, C) = 0$
for the (non-)SUSY, (non-)interacting and (non-)relativistic systems of particles which are found to be reparameterization 
(i.e. 1D diffeomorphism) invariant. We have noted earlier that, as far as the gauge symmetry transformations and corresponding 
(anti-)BRST symmetry transformations are concerned, the CF-type restriction is {\it trivial} for {\it all} the systems  that have been 
considered in our present endeavor except the free massive spinning (i.e. SUSY) relativistic particle. However, 
the CF-type restriction:  $B + \bar B + i\, (\bar C\, \dot C - \dot {\bar C}\, C) = 0$
is {\it universal} for {\it all} the 1D diffeomorphism (i.e. reparameterization) invariant models that have 
been discussed in  our present research-cum-review article.

To derive the (anti-)BRST symmetry transformations for the phase variables ($x_\mu,\, p^\mu$) and the CF-type restriction for {\it all}
the 1D diffeomorphism (i.e. reparameterization) invariant systems that have been considered in our present endeavor, 
first of all, we generalize the {\it basic} infinitesimal diffeomorphism transformations: 
$\tau \longrightarrow \tau - \varepsilon (\tau)$ (that are defined on the 1D {\it ordinary} spacetime manifold) 
onto a (1, 2)-dimensional supermanifold, as [7, 8]
\begin{eqnarray}
\tau - \varepsilon (\tau) \longrightarrow \tau - \theta\,\bar C (\tau) - \bar\theta\, C(\tau)+ \theta\, \bar\theta\, f (\tau),
\end{eqnarray}
where the fermionic ($\bar C^2 = C^2 = 0,\; \bar C\, C + C\, \bar C = 0$) (anti-)ghost variables 
are the coefficients of a pair of Grassmannian variables ($\theta, \, \bar\theta$) that characterizes the (1, 2)-dimensional 
supermanifold  (along with a bosonic variable $\tau$). The secondary variable $f(\tau)$ in (66) is to be  
determined  by the consistency condition. The {\it latter} is nothing but the requirement of the property of 
absolute anticommutativity of the (anti-)BRST symmetry transformations which is one of the sacrosanct 
properties of the  (anti-)BRST symmetry transformations. Let us, first of all, focus 
on the coordinates $x^\mu (\tau)$ and find out the {\it quantum} (anti-)BRST symmetry transformations (corresponding
to its {\it classical} diffeomorphism symmetry transformation: $\delta_r x_\mu  = \varepsilon \, \dot x^\mu$). 
For this purpose, we generalize 
\begin{eqnarray}
x^\mu (\tau) \;\longrightarrow \; X^\mu (\tau, \theta, \bar\theta) \; \longrightarrow \; \tilde X^\mu (\tau -\theta\, \bar C 
- \bar\theta\, C + \theta\, \bar\theta\, f(\tau), \,  \theta, \, \bar\theta).
\end{eqnarray}
Thus, we note  that, it is a {\it two} steps of generalizations that we require for the derivation
of the (anti-)BRST symmetry transformations and CF-type restriction when  the diffeomorphism symmetry transformation(s) are 
present in a theory. This should be contrasted with the gauge symmetry transformations where, within the framework of Bonora-Tonin 
superfield/supervariable approach [9-11], only {\it one} step is required. In other words, in the context of the {\it classical}
gauge symmetry, if we wish to compute the corresponding {\it quantum} (anti-)BRST symmetry transformations (within the 
framework of Bonora-Tonin supervariable/superfield approach [9-11]), we have to take into account
\begin{eqnarray}
x_\mu (\tau) \;\longrightarrow \; X_\mu (\tau, \theta, \bar\theta) & = & x_\mu (\tau) + \theta\, \bar R_\mu (\tau) + 
\bar\theta\,R_\mu (\tau) + \theta\, \bar\theta\, S_\mu (\tau)\nonumber\\
& \equiv  & x_\mu (\tau) + \theta\, (s_{ab} x_\mu) +  \bar\theta\, (s_{b} x_\mu) + \theta\, \bar\theta\,  (s_b\,s_{ab} x_\mu),
\end{eqnarray}
where, on the r.h.s., the fermionic secondary variables $R_\mu (\tau)$ and $\bar R_\mu (\tau)$ and a bosonic variable $S_\mu (\tau)$
have to be determined so that one can obtain the (anti-)BRST symmetry transformations $s_{(a)b}$. The establishment of relationships:
$s_b\longleftrightarrow \partial_{\bar\theta}\, |_{\theta = 0}$ and $s_{ab}\longleftrightarrow \partial_{\theta}\, |_{\bar\theta = 0}$
is one of the highlights of a set of very nicely written papers in Refs. [9-11].

Let us focus on  $\tilde X^\mu (\tau -\theta\, \bar C - \bar\theta\, C + \theta\, \bar\theta\, f, \,  \theta, \, \bar\theta)$
vis-\`a-vis the super expansion (68) on the (1, 2)-dimensional super manifold.  The analogue of (68) can be written as 
\begin{eqnarray}
&& \tilde X^\mu (\tau -\theta\, \bar C - \bar\theta\, C + \theta\, \bar\theta\, f, \,  \theta, \, \bar\theta)\nonumber\\
 & = &  {\cal X^\mu} (\tau -\theta\, \bar C - \bar\theta\, C + \theta\, \bar\theta\, f) 
 +  \theta\, \bar R^\mu (\tau -\theta\, \bar C - \bar\theta\, C + \theta\, \bar\theta\, f)\nonumber\\
& + &  \bar\theta\, R^\mu (\tau -\theta\, \bar C - \bar\theta\, C + \theta\, \bar\theta\, f)
 +  \theta\,\bar\theta\, S^\mu (\tau -\theta\, \bar C - \bar\theta\, C + \theta\, \bar\theta\, f),
\end{eqnarray}
where we note that all the secondary super variables on the r.h.s. are {\it still} function of the {\it general} infinitesimal 
super diffeomorphism symmetry transformations (66). However, as we have observed, the connection of the above super expansion 
with the (anti-)BRST transformations $s_{(a)b}$ can be established if and only if we go from the super expansion (69)
to the super expansion (68) by exploiting the mathematical tricks  of the Taylor expansion. In this context, we note the following 
(with inputs: $\theta^2 = \bar\theta^2 = 0,\; \theta\,\bar\theta + \bar\theta \,\theta = 0$), namely;
\begin{eqnarray}
\theta\,\bar\theta\,S_\mu(\tau - \theta\,\bar C - \bar\theta\,C + \theta\,\bar\theta\,f) &=& \theta\,\bar\theta\,S_\mu(\tau), \nonumber\\
\theta\,\bar R_\mu (\tau - \theta\,\bar C - \bar\theta\,C + \theta\,\bar\theta\,f) &=& \theta\,\bar R_\mu(\tau) - \theta\,\bar\theta \,
C\,\dot{\bar R}_\mu(\tau),\nonumber\\
\bar\theta\,R_\mu(\tau - \theta\,\bar C - \bar\theta\,C + \theta\,\bar\theta\,f) 
& = & \bar\theta\,R_\mu (\tau) + \theta\,\bar\theta \, \bar C\,\dot R_\mu (\tau),\nonumber\\
{\cal X}^\mu (\tau - \theta\,\bar C - \bar\theta\,C + \theta\,\bar\theta\,f) & = & {x^\mu}(\tau) 
- \theta\,\bar C\,\dot x^\mu(\tau) - \bar\theta\,C\,\dot x^\mu(\tau)\nonumber\\ 
&+& \theta\,\bar\theta\, \big [f\,\dot x^\mu (\tau) - \bar C  \, C \,\ddot x^\mu (\tau) \big ],
\end{eqnarray} 
which demonstrate that, on the r.h.s. of the above expansions, we have obtained the secondary  
variables as function of $\tau$ {\it only}. Now collecting all the coefficients of $\theta, \, \bar\theta$ and $\theta\, \bar\theta$
from (70), we obtain the following explicit expansion:  
\begin{eqnarray}
&& \tilde X_\mu (\tau -\theta\, \bar C - \bar\theta\, C + \theta\, \bar\theta\, f, \,  \theta, \, \bar\theta)
 =  {x_\mu} (\tau) +  \theta\, (\bar R_\mu -\bar C \, \dot x_\mu)\nonumber\\
& + &  \bar\theta\, (R_\mu -  C \dot x_\mu)
 +  \theta\,\bar\theta\, [ S_\mu +  \bar C\, \dot R_\mu  - C \, \dot R_\mu  + f\, \dot x_\mu - \bar C\, C\, \ddot x_\mu]. 
\end{eqnarray}
At this juncture, we exploit the theoretical strength of the HC (in the context of our present set of 1D diffeomorphism invariant 
theories) which is nothing but the requirement that a Lorentz scalar does {\it not} transform under any kind of internal, spacetime, 
supersymmetric, diffeomorphism, etc., transformations. Hence, we have the following restriction:    
\begin{eqnarray}
\tilde X_\mu (\tau -\theta\, \bar C - \bar\theta\, C + \theta\, \bar\theta\, f, \,  \theta, \, \bar\theta) 
= X_\mu (\tau, \theta, \, \bar\theta) \equiv {x_\mu} (\tau).
\end{eqnarray}
In other words, the  whole expansion on the r.h.s. is equal to $x_\mu(\tau)$. Thus, we have the following equality
due to the application of HC, namely; 
\begin{eqnarray}
 {x_\mu} +  \theta\, (\bar R_\mu -\bar C \, \dot x_\mu)
 +   \bar\theta\, (R_\mu -  C \dot x_\mu)
 +  \theta\,\bar\theta\, [ S_\mu +  \bar C\, \dot R_\mu  - C \, \dot R_\mu  + f\, \dot x_\mu - \bar C\, C\, \ddot x_\mu] = x_\mu.
\end{eqnarray}
As a consequence, we obtain the following relationships between the secondary variables $R_\mu (\tau), \, \bar R_\mu (\tau), \, 
S_\mu (\tau)$ and the basic variables of the BRST invariant theory, namely; 
\begin{eqnarray}
R_\mu  =   C \dot x_\mu, \qquad \bar R_\mu  = \bar C \, \dot x_\mu, \qquad 
S_\mu =   C \, \dot R_\mu   - \bar C\, \dot R_\mu   -  f\, \dot x_\mu + \bar C\, C\, \ddot x_\mu. 
\end{eqnarray}
The substitutions of $R_\mu$ and $\bar R_\mu$ into $S_\mu$ yield the following 
\begin{eqnarray}
S_\mu =   -\, [(\dot {\bar C}\, C + \bar C\, \dot C + f)\, \dot x_\mu + \bar C\, C \, \ddot x_\mu]. 
\end{eqnarray}
As a result, we have obtained  {\it all} the secondary variables $R_\mu(\tau), \, \bar R_\mu(\tau)$ and $  S_\mu(\tau)$
in terms of the basic variables of the (anti-)BRST invariant theory and $f(\tau)$.

Plugging in the values of $R_\mu, \, \bar R_\mu$ and $  S_\mu$ in our earlier equation (68), we observe the following
[due to $\tilde X_\mu (\tau -\theta\, \bar C - \bar\theta\, C + \theta\, \bar\theta\, f, \, 
 \theta, \, \bar\theta)  = X_\mu (\tau, \theta, \, \bar\theta)$], namely; 
\begin{eqnarray}
X_\mu (\tau, \theta, \, \bar\theta) &=&  {x_\mu} (\tau) + \theta \,  (\bar C \, \dot x_\mu) + \bar\theta \, ( C \, \dot x_\mu)
+ \theta\bar\theta \, [-\, (\dot {\bar C}\, C + \bar C\, \dot C + f)\, \dot x_\mu + \bar C\, C \, \ddot x_\mu] \nonumber\\
&\equiv & {x_\mu} (\tau) + \theta \, (s_{ab}\, x_\mu) + \bar\theta \, (s_{b}\, x_\mu) + \theta\bar\theta\, (s_b\, s_{ab}\, x_\mu),
\end{eqnarray}
where we have utilized our observation in (72). A close look at the coefficient of $\theta \, \bar\theta $ shows that we have the 
presence of the secondary variable $f(\tau)$ of equation (66) in {\it it}. To derive this variable in terms of the basic and 
auxiliary variables, we require the validity of the 
absolute anticommutativity property. In other words, we have the following
\begin{eqnarray}
s_b\, s_{ab}\, x_\mu = -\, s_{ab}\,  s_b\, x_\mu \quad \Longrightarrow  \quad \{s_b, \, s_{ab}\}\, x_\mu = 0, 
\end{eqnarray}
where we have to use: $s_b\, x_\mu = C \, \dot x_\mu, \, s_{ab}\, x_\mu = \bar C \, \dot x_\mu$. 
Writing the l.h.s. and r.h.s. of the above equation  explicitly, we obtain
\begin{eqnarray}
s_b\, s_{ab}\, x_\mu  = (i\, B - \bar C\, \dot C)\, \dot x_\mu - \bar C\, C\, \ddot x_\mu, \nonumber\\
-\, s_{ab}\,  s_b\, x_\mu = (-\, i\, \bar B - \dot{\bar C}\, C)\, \dot x_\mu - \bar C\, C\, \ddot x_\mu, 
\end{eqnarray}
which are separately and independently equal to $S_\mu$ of equation (75). 
It is worthwhile to point out that,  in (78), we have used the standard symmetry transformations of BRST formalism: 
$s_b\, \bar C = i\, B$ and $s_b\, C = i\, \bar B$. This equality leads to the following  relationship
\begin{eqnarray}
f\, (\tau) = i\, \bar B - \bar C\, \dot C \equiv -\, i\, B - \dot{\bar C}\, C \Longrightarrow 
B + \bar B + i\,(\bar C\,\dot C - \dot{\bar C}\,C) = 0.
\end{eqnarray}
which is nothing but the CF-type restriction on our 1D diffeomorphism invariant theory.

We wrap-up this section with the following {\it final} remarks. First, the MBTSA to BRST formalism is capable of providing the 
precise derivations of the (anti-)BRST symmetry transformations: $s_b x_\mu  = C\, \dot x_\mu, \, s_b p_\mu  =
 C\, \dot p_\mu, \, s_{ab} x_\mu  = \bar C\, \dot x_\mu, \, s_{ab} p_\mu  = \bar C\, \dot p_\mu$ and the (anti-) BRST 
 invariant CF-type restriction: $B + \bar B + i\,(\bar C\,\dot C - \dot{\bar C}\,C) = 0$. However, it does not shed any light on 
 the (anti-)BRST symmetry transformations corresponding to the einbein variable $e(\tau)$.
 Second, the (anti-)chiral supervariable approach (ACSA) to BRST formalism (see, e.g. [50, 51] and references therein) has been exploited for the 
 derivations of the (anti-) BRST symmetries for the {\it rest} of the variables of the theory in our earlier works [30-34]
 {\it besides} the phase variables $(x_\mu,\, p^\mu)$ for which the MBTSA has been exploited. 
Third, the standard (anti-)BRST symmetry transformations: $s_b \bar C = i\, B, \,s_b B = 0, \,s_{ab} C = i\, \bar B, 
\,s_{ab} \bar B = 0$ have been taken into account in our present endeavor and earlier works [30-34] {\it without} any precise derivation.
Finally, in our Appendix C, we have derived the (anti-)BRST symmetry transformations for the {\it other} variables of the present theory 
by exploiting the (i) property of the off-shell nilpotency, (ii) property of the absolute anticommutativity, and (iii) 
invariance of the CF-type restriction under the (anti-)BRST symmetry transformations.

\section{ Nilpotent (Anti-)BRST Symmetries and CF-Type Restrictions: 2D Diffeomorphism Invariant Model of a Bosonic String Theory}

The 2D diffeomorphism invariant models of the bosonic strings  and string field 
theories\footnote{Using the light-cone gauge in the case of the Polyakov Lagrangian density, a bosonic string theory is BRST quantized [52] where the 
emphasis is laid on the conformal symmetry and conformal anomaly. The {\it latter} has been shown to be {\it zero} in $D = 26$. 
On the other hand, a new BRST quantization procedure [53] has been adopted for the quantization of the open bosonic string theory. 
As far as the bosonic string field theory is concerned, the BRST quantization of Witten's bosonic field theory has been performed in [55]. 
The bosonic string and associated field theory have {\it also} been studied by others (see, e.g. [54, 56] and references therein). 
However, none of the above approaches has been able to lay  emphasis on the existence of the CF-type restrictions corresponding to the BRST formulation of
the  {\it classical}  infinitesimal 2D diffeomorphism (i.e. world-sheet) symmetry transformations. We have achieved this goal
 for the BRST quantized bosonic string theory  [32] proposed by 
Kato and Ogawa (see, e.g. [6, 32] for details). } (see, e.g. [52-56] and references therein) 
are the first few steps for the  generalizations of the 1D diffeomorphism invariant models of the non-SUSY relativistic particles. 
These generalizations  have,
ultimately, led to their {\it further} generalizations to  the ideas of superstring theories (see, e.g. [1-5] for details). The {\it latter} theories
are the most promising candidates to provide (i) the complete unification of all the four fundamental interactions
of nature, and (ii) a consistent theory of quantum gravity. In our present section, we focus on a specific model of the 2D 
diffeomorphism invariant bosonic string theory  whose classical, infinitesimal and continuous diffeomorphism symmetry
transformations were elevated to their {\it quantum} versions of the BRST symmetry transformations by Kato and Ogava [6].

Unlike the previous sections where we have focused on the 1D diffeomorphism (i.e. reparameterization) invariant as well as the gauge
invariant models of the (non-)interacting, (non-)SUSY and (non-)relativistic particles, we concentrate, in our 
present section, on the 2D diffeomorphism symmetry invariant model of a bosonic string theory (purposed by Kato and Ogawa [6])
where there is {\it no} gauge invariance because {\it this} 2D diffeomorphism invariant theory is {\it not} endowed with the first-class constraints in
the terminology of Dirac's prescription for the classification scheme of constraints [22, 23]. 
Here the 2D diffeomorphism symmetry transformation is present on the 2D world-sheet that is traced out by the propagation  of 
a bosonic string which is embedded in a target space which is a D-dimensional $(D > 2)$ flat Minkowskian spacetime manifold. We begin with 
the following Lagrangian density for the bosonic string with $k$ as the string tension parameter [6]
\begin{eqnarray}
{\cal L}_0 = - \frac{1}{2k}\, {\tilde g}^{ab}\,\partial_a \,X^\mu \, \partial_b \, X_\mu 
+ E\, (\mbox{det} \,\tilde g + 1),
\end{eqnarray}
where ${\tilde g}^{ab} = \sqrt{- g} \,g^{ab}$ has {\it two} independent degrees of freedom due to the EoM with respect 
to the Lagrange multiplier field  $E$ which yields: $\mbox{det} \,{\tilde g} = - 1 $. The choice:
${\tilde g}^{ab} = \sqrt{- g} \,g^{ab}$ has been taken into account to avoid the conformal anomaly that arises when 
one regularizes the system in a gauge invariant manner (see, e.g. [57, 58] for details). 
It is worth pointing out that the starting  Lagrangian density 
$\tilde {\cal L}_0 = - \frac{1}{2k}\, \sqrt{- g} \,g^{ab}\,\partial_a \,X^\mu \, \partial_b \, X_\mu $
contains  the inverse ($g^{ab}$) of the original 2D metric tensor $g_{ab}$ whose determinant is: $g = $ det $g_{ab}$.
The 2D surface, traced out by the propagation   
of the bosonic string, is parameterized by $\xi^a = (\xi^0, \,\xi^1) \equiv (\tau,\, \sigma)$ where 
$- \, \infty  <  \tau < +\, \infty $ is the evolution parameter and $\sigma = 0, \, \pi$ defines the length of 
the string. In the above Lagrangian density, the fields $X^\mu (\xi)$ (with $\mu = 0, 1, 2, 3, ..., D-1$)
are the string coordinates which are Lorentz vectors in the D-dimensional flat Minkowskian target space 
{\it but} are scalars w.r.t. the 2D surface traced out by the propagation of the bosonic string that is parameterized by 
$\xi^a = (\xi^0, \,\xi^1) \equiv (\tau,\, \sigma)$ and the symbols
$\partial_a = (\partial_0, \, \partial_1) \equiv (\partial_\tau, \, \partial_\sigma)$ are the partial derivatives that are defined on the 2D surface. 
It can be explicitly checked that the following infinitesimal, continuous and {\it classical} 2D diffeomorphism symmetry transformations 
$(\delta_\varepsilon )$, namely; 
\begin{eqnarray}
&& \delta_\varepsilon \, X^\mu  = \varepsilon^a \,\partial_a X^\mu , \qquad 
\delta_\varepsilon \, E = \partial_a (\varepsilon^a \, E), 
\qquad \delta_\varepsilon \,(\mbox{det} \,{\tilde g}) = \varepsilon^a \,\partial_a (\mbox{det}\, {\tilde g} ),\nonumber\\
&& \delta_\varepsilon \,{\tilde g}^{ab} = \partial_m (\varepsilon^m \, {\tilde g}^{ab}) \, - \, 
(\partial_m \varepsilon^a) \,{\tilde g}^{mb} \, - \, (\partial_m \, \varepsilon^b)\, {\tilde g}^{am}, 
\end{eqnarray}
lead to $\delta_\varepsilon \, {\cal L}_0 =  \partial_a \, (\varepsilon^a \,{\cal L}_0)$ which renders the action integral 
$S = \int d^2 \xi \, {\cal L}_0 \equiv \int_{- \, \infty}^{+ \, \infty} d\tau \int_{0}^{\pi} d\sigma\, {\cal L}_0$
invariant provided the infinitesimal diffeomorphism transformation parameter [$\varepsilon^a (\xi)$]
satisfy the proper boundary conditions at  $\sigma = 0$ and $\sigma = \pi$. Here the basic diffeomorphism symmetry transformations on the 
2D surface are: $\xi^a \to \xi^{{a}{'}} \equiv  \xi^a - \varepsilon^a (\xi)$ where $\varepsilon^a (\xi)$ are the 
2D infinitesimal diffeomorphism transformation parameters. In this transformation, we have chosen the Latin indices:
$a, b, c, ..., m, n, ... = 0, 1$ which denote the $\tau$ and $\sigma$ directions on the 2D surface 
(that is  traced out by the propagation of the bosonic string).

Using the standard BRST prescription for the BRST symmetry transformations ($s_b$), we replace the infinitesimal 2D 
diffeomorphism parameter $\varepsilon ^a$ by the 2D fermionic [$(C^a)^2 = 0, \, C^a\, C^b + C^b\, C^a = 0$] ghost fields $C^a$. 
Thus, we obtain the following infinitesimal, continuous and off-shell nilpotent $(s_b^2 = 0)$ BRST symmetry transformations $(s_b)$ as  
\begin{eqnarray}
&&s_b X^\mu = C^a \, \partial_a \, X^\mu, \qquad  s_b\, E = \partial_a \,(C^a E),\qquad  
s_b(\mbox{det} \, {\tilde g}) = \varepsilon^a \, \partial_a \, (\mbox{det} {\tilde g}), \nonumber\\
&&s_b C^a = C^m \, \partial_m \, C^a,\,\,\qquad s_b \bar C^a = i B^a,\qquad  s_b \,B^a = 0,\nonumber\\
&& s_b\,{\tilde g}^{ab} = \partial_m (C^m \, {\tilde g}^{ab}) \, - \, 
(\partial_m\,C^a) \,{\tilde g}^{mb} \, - \, (\partial_m \, C^b)\, {\tilde g}^{am}, 
\end{eqnarray}
where $B^a = (B^0, \, B^1)$ are the Nakanishi-Lautrup type  auxiliary fields, $\bar C^a \, (a = 0, 1)$ 
are the anti-ghost fields and we have obtained $s_b\, C^a = C^b\, \partial_b\, C^a$ by the requirement of the 
off-shell nilpotency ($s_b^2\, X^\mu = 0$) property. To BRST-quantize the theory, we have to gauge-fix the Lagrangian density and incorporate the
Faddeev-Popov (FP) ghost terms. For this purpose, first of all, we decompose  the modified version of the inverse of the  metric
 ${\tilde g}^{ab}$ in its component forms as follows (see, e.g. [6] for details)
\begin{eqnarray}
{\tilde g}^{ab} = 
\begin{pmatrix} 
A_1 + A_2  & A_0  \\ A_0 & A_1 - A_2 
\end{pmatrix},
\end{eqnarray}
and take $A_0 = A_1 = 0$ so that  $\det \,{\tilde g} = - 1$ yields ${A_2}^2 = 1$.
It is obvious that we obtain the {\it flat} limit from the inverse of the modified metric 
 ${\tilde g}^{ab} \Rightarrow  \eta^{ab} \equiv (+1, \, -1)$ when we make the choice $A_2 = + 1$.
Using the standard technique of the BRST formalism, we add the gauge-fixing and FP-ghost terms 
to the Lagrangian density ${\cal L}_0$ which are [59, 33, 34]
\begin{eqnarray}
{\cal L}_{gf} + {\cal L}_{FP} = s_b \,\big[-\,i\,\bar C_0 A_0 - i\,\bar C_1 A_1 \big],
\end{eqnarray}
to yield the Lagrangian density ${\cal L}_b$.  It will be worthwhile to mention here that there is a more general method to obtain the gauge-fixing
and FP-ghost terms which has been {\it recently} proposed in an interesting piece of work [60]. This theoretical methodology leads to the following explicit 
expression for the BRST-invariant Lagrangian density 
\begin{eqnarray}
{\cal L}_b &=& {\cal L}_0 + B_0 A_0 + B_1 A_1 + i\, \big[C^a\partial_a \bar C_0 - \bar C_0 (\partial_a C^a)
- \bar C_1 (\partial_0 C^1 + \partial_1 C^0)\big]\, A_0 \nonumber\\
&+& i \,\big[ C^a \,\partial_a \,\bar C_1 - \bar C_1 \,(\partial_a \, C^a) -
 \bar C_0 (\partial_0 C^1 + \partial_1 C^0)\big] \,A_1 \nonumber\\ 
&-& i \,\big[\bar C_0 \,(\partial_0 \, C^1\, - \partial_1 \, C^0) + \bar C_1\, (\partial_0 C^0\, - \partial_1 C^1)\big]\, A_2, 
\end{eqnarray}
where we have used the following BRST symmetry transformations for the component gauge fields ($A_0, \, A_1, \, A_2$)
from the transformation $s_b\, {\tilde g}^{ab}$ [cf. Eq. (82)], namely;
\begin{eqnarray}
&&s_b A_0 = C^a\,\partial_a\, A_0 - (\partial_0 \, C^1 + \partial_1 \, C^0)\,  A_1 - 
(\partial_0 \, C^1 - \partial_1 \, C^0) \, A_2, \\ \nonumber
&&s_b A_1 = C^a\,\partial_a \, A_1 - (\partial_0 \, C^0 - \partial_1 \, C^1) \, A_2 - 
(\partial_1 \, C^0 + \partial_0 \,  C^1)\, A_0, \\ \nonumber
&&s_b A_2 = C^a\,\partial_a \, A_2 - (\partial_0 \, C^0 - \partial_1 \, C^1) A_1 - 
(\partial_1 \, C^0 - \partial_0 \, C^1) A_0. 
\end{eqnarray}
As is obvious, the (anti-)ghost fields $(\bar C^a)C^a$ carry the ghost numbers $(-1)+1$, respectively. It is straightforward to check that 
${\cal L}_b$ transforms to a total derivative on the 2D surface under the BRST symmetry transformations $(s_b)$, namely; 
\begin{eqnarray}
s_b\, {\cal L}_{b} &=& \partial_a \, \Bigl[C^a \big({\cal L}_{0} + B_0\,A_0 + B_1\,A_1\big) + 
i\, \bar C_1 \, C^a (\partial_0 \, C^1 + \partial_1 \, C^0) \, A_0 \nonumber\\ &+& i\, \bar C_0 C^b\,\partial_b \, (C^a\,A_0)
+ i\,\bar C_0 \, C^a \,(\partial_0 \, C^1 + \partial_1 \, C^0) \, A_1 
+ i\, \bar C_1 C^b\,\partial_b \, (C^a\,A_1)\nonumber\\
&+& i\,\bar C_0 \, C^a \,(\partial_0 \, C^1 - \partial_1 \, C^0) \, A_2 +  
i\,\bar C_1 \, C^a \,(\partial_0 \, C^0 - \partial_1 \, C^1) \, A_2\Bigr]. 
\end{eqnarray}
As a consequence, it is obvious  that the action integral $ S = \int_{-\infty}^{+ \infty} d \tau \int_0^\pi\, d \sigma \, {\cal L}_b$ 
remains invariant for the physically well-defined fields which vanish off at appropriate boundary conditions at $\sigma = 0$ and $\pi$ as well as $\tau \rightarrow \pm \infty$.

Against the backdrop of the above discussions on the BRST symmetry transformations, we now
focus on the corresponding anti-BRST symmetry transformations which have been derived in our earlier work [33]. 
These off-shell nilpotent ($ s_{ab}^2 = 0$) anti-BRST symmetry transformations are as follows 
\begin{eqnarray}
&& s_{ab}\, X^\mu = \bar C^a \, \partial_a X^\mu, \quad \qquad  s_{ab} \,\bar C^a = \bar C^m \, \partial_m \,\bar C^a, \qquad 
\quad  s_{ab}\, C^a = i\, \bar B^a, \nonumber\\
&&  s_{ab} \,E = \partial_a \, (\bar C^a \,E), \qquad  s_{ab} \, (\mbox{det} {\tilde g}) = 
\bar C^a \, \partial_a (\mbox{det} {\tilde g}), \qquad s_{ab}\, \bar B^a = 0,\nonumber\\
&&s_{ab}\, {\tilde g}^{ab} = \partial_m \,(\bar C^m\, {\tilde g}^{ab}) - (\partial_m \,\bar C^a) \,{\tilde g}^{mb}
-  (\partial_m \,\bar C^b) \,{\tilde g}^{am},
\end{eqnarray}
where a new Nakanishi-Lautrup type auxiliary field $\bar B^a\, (\xi )$ has been invoked for the sake of generality. 
We further note that the anti-BRST symmetry transformations for the component fields $(A_0, \, A_1, \, A_2)$ can be derived from 
$s_{ab}\, {\tilde g}^{ab}\, (\xi)$ [cf. Eq. (88)] as follows:
\begin{eqnarray}
&& s_{ab} A_0 = \bar C^a\,\partial_a\, A_0 - (\partial_0 \, \bar C^1 + \partial_1 \, \bar C^0)\,  A_1 - 
(\partial_0 \, \bar C^1 - \partial_1 \,\bar C^0) \, A_2, \nonumber\\
&& s_{ab} A_1 = \bar C^a\,\partial_a \, A_1 - (\partial_0 \, \bar C^0 - \partial_1 \, \bar C^1) \, A_2 - 
(\partial_1 \,\bar C^0 + \partial_0 \, \bar C^1)\, A_0,  \nonumber\\
&& s_{ab} A_2 = \bar C^a\,\partial_a \, A_2 - (\partial_0 \, \bar C^0 - \partial_1 \, \bar C^1) A_1 - 
(\partial_1 \, \bar C^0 - \partial_0 \, \bar C^1) A_0.
\end{eqnarray}
It goes without saying that the anti-BRST symmetry transformations (88) and (89) are the analogues of the equations (82) and (86)
for the BRST symmetry transformations.

With the help of the anti-BRST symmetry transformations (88) and (89), we can find out the gauge-fixing and FP-ghost terms for the {\it perfectly} 
anti-BRST invariant Lagrangian density by following the standard BRST technique, as [59, 33]
\begin{eqnarray}
{\cal L}_{gf} + {\cal L}_{FP} = s_{ab}\,\big[i\,C_0 A_0 +  i\, C_1 A_1 \big],
\end{eqnarray}
which leads to the total anti-BRST invariant Lagrangian density (${\cal L}_{\bar b}$) as 
\begin{eqnarray}
{\cal L}_{\bar b} & = &{\cal L}_0 - {\bar B}_0 A_0 - \bar B_1 A_1 
+ i\,\Bigl [C_0 (\partial_a \, \bar C^a) + (\partial_a C_0)\, \bar C^a 
+ C_1 \, (\partial_0 \, \bar C^1 + \partial_1 \, \bar C^0)\Bigr] \, A_0 \nonumber\\
&+& i\,\Bigl [C_1 \, (\partial_a \, \bar C^a) + (\partial_a \, C_1) \,\bar C^a 
+ C_0 \, (\partial_0 \, \bar C^1 + \partial_1 \, \bar C^0)\Bigr] \, A_1  \nonumber\\
& + & i\,\Bigl[C_0 \, (\partial_0 \, \bar C^1 - \partial_1 \, \bar C^0) 
+ C_1 \,(\partial_0 \, \bar C^0 - \partial_1 \, \bar C^1) \Bigr] \, A_2,
\end{eqnarray}
where some total derivative terms have been dropped because they do {\it not} contribute anything in the description of the dynamics of our theory.  
The application of the anti-BRST symmetry transformations ($s_{ab}$) on the Lagrangian density ${\cal L}_{\bar b}$ leads to the following:
\begin{eqnarray}
s_{ab}\, {\cal L}_{\bar b} &=& \partial_a \, \Bigl[\bar C^a \big({\cal L}_{0} - \bar B_0\,A_0 - \bar B_1\,A_1\big) - 
i\, C_1 \, \bar C^a (\partial_0 \, \bar C^1 + \partial_1 \, \bar C^0) \, A_0 \nonumber\\
&-& i\, C_0 \,\bar C^b\,\partial_b \, (\bar C^a\,A_0)
- i\,C_0 \, \bar C^a \,(\partial_0 \, \bar C^1 + \partial_1 \, \bar C^0) \, A_1 
- i\,  C_1\,\bar C^b\,\partial_b \, (\bar C^a\,A_1)\nonumber\\
&-& i\, C_0 \, \bar C^a \,(\partial_0 \, \bar C^1 - \partial_1 \, \bar C^0) \, A_2 -  
i\, C_1 \, \bar C^a \,(\partial_0 \, \bar C^0 - \partial_1 \, \bar C^1) \, A_2\Bigr]. 
\end{eqnarray}
As a consequence, it is clear that action integral  $ S = \int_{-\infty}^{+ \infty} d \tau \int_0^\pi\, d \sigma \, {\cal L}_{\bar b}$ 
will remain invariant under the anti-BRST symmetry transformations for the physical fields that vanish off when we take the 
limit $\tau \longrightarrow \pm \infty$ and $\sigma = 0, \pi$.

We conclude this section with the following decisive  and useful remarks. First of all, we note that, in the BRST analysis of 
our present bosonic string, {\it only} the BRST symmetry transformations have been quoted by 
Kato and Ogawa [6]. Second, the appropriate anti-BRST symmetry transformations for {\it this} theory have been obtained   
in our earlier work (see, e.g. [33] for details). Third, there is an overall sign difference between our BRST symmetry 
transformations and that of Kato and Ogawa because our basic diffeomorphism symmetry transformations
($\xi^a \to \xi^{{a}{'}} \equiv  \xi^a - \varepsilon^a (\xi)$) are different from the {\it former} by a sign factor.
Fourth, the requirement of the absolute anticommutativity (i.e. $\{s_b, s_{ab} \} = 0$) between the BRST and anti-BRST symmetry
transformations leads to the following 
\begin{eqnarray}
\{s_b, s_{ab}\}\,X^\mu &=&  i\,\Big[B^a + \bar B^a + i\, (C^m \, \partial_m \, \bar C^a + \bar C^m \, \partial_m \, C^a)\Big]\,\big(\partial_a\,
X^\mu\big) = 0, \nonumber\\
\{s_b, s_{ab}\}\,E &=& i\,\partial_a\,\Big[\{B^a + \bar B^a + i\, (C^m \, \partial_m \, \bar C^a + \bar C^m \, \partial_m \, C^a)\}\,E\Big] = 0, \nonumber\\
\{s_b, s_{ab}\}\,\tilde g^{mn} &=& i\,\partial_k\,\Big[\big\{B^k + \bar B^k + i\, (C^l \, \partial_l \, \bar C^k + \bar C^l \, \partial_l \, C^k)
\big\}\,\tilde g^{mn}\Big] \nonumber\\
&-& i\,\partial_k\,\Big[B^m + \bar B^m + i\, (C^l \, \partial_l \, \bar C^m + \bar C^l \, \partial_l \, C^m)\Big]\,\tilde g^{kn} \nonumber\\
&-& i\,\partial_k\,\Big[B^n + \bar B^n + i\, (C^l \, \partial_l \, \bar C^n + \bar C^l \, \partial_l \, C^n)\Big]\,\tilde g^{km} = 0,
\end{eqnarray}
which demonstrates that the CF-type restrictions:  $B^a + \bar B^a + i \; (\bar C^m \;\partial_m \;C^a 
+ C^m \; \partial_m \;\bar C^a) = 0$ have to be invoked for the validity of $\{s_b, s_{ab} \} = 0$.
Fifth, we observe that the CF-type restrictions are (anti-)BRST invariant: 
 $s_{(a)b}\; [B^m + \bar B^m + i \; (\bar C^n \;\partial_n \;C^m  + C^n\; \partial_n \;\bar C^m)]  = 0$
which demonstrates that these restrictions are {\it physical} and essential for the proof of independent identity of 
$s_b$ and $s_{ab}$. Finally, we point out  that the requirement of the absolute anticommutativity property on the 
(anti-)ghost fields $(\bar C^a)C^a$ leads to the (anti-) BRST symmetry transformations on the Nakanishi-Lautrup auxiliary fields ($B^a, \, \bar B^a$) as:
\begin{eqnarray}
\{s_b, s_{ab}\}\,\bar C^a  =  0 \Longrightarrow s_{b}\, \bar B^a =   C^m\, \partial_m\, \bar B^a - \bar B^m\, \partial_m\,  C^a,  \nonumber\\
\{s_b, s_{ab}\}\,C^a  =  0 \Longrightarrow s_{ab}\, B^a =  \bar C^m\, \partial_m\, B^a - B^m\, \partial_m\, \bar C^a. 
\end{eqnarray}
It is straightforward to check that $s_{ab}^2 \, B^a = 0$ and $ s_b^2\, \bar B^a = 0$ which show that the  (anti-)BRST symmetry transformations
of Nakanishi-Lautrup auxiliary fields ($B^a, \, \bar B^a$) are {\it proper} and they satisfy the absolute 
anticommutativity property (i. e. $\{s_b, s_{ab}\}\,B^a = 0, \; \{s_b, s_{ab}\}\,\bar B^a = 0 $), too. 
In the proof of the (anti-)BRST invariance (i. e. $s_{(a)b}\, [B^a + \bar B^a + i \; (\bar C^m \;\partial_m \;C^a 
+ C^m \; \partial_m \;\bar C^a)] = 0$), we have to make use of (82), (88) and the validity of the CF-type restrictions: 
$B^a + \bar B^a + i \; (\bar C^m \;\partial_m \;C^a + C^m \; \partial_m \;\bar C^a)] = 0$ (with $a, \, m = 0, 1$).\\


\section{CF-Type Restrictions for the 2D Diffeomorphism Invariant Theory: MBTSA to BRST Formalism}

To derive the 2D version of the {\it universal} D-dimensional CF-type restrictions
(see, e.g. [8] for details), we note that the 2D infinitesimal diffeomorphism 
[$\xi^a \rightarrow   \xi^a - \varepsilon^a (\xi), \, a = 0, 1$] symmetry transformations have to be generalized 
as the infinitesimal super diffeomorphism symmetry transformations, onto 
a suitably chosen (2,2)-dimensional supermanifold, as [34]
\begin{eqnarray}
\xi^a \longrightarrow   \xi^a -  \theta\, \bar C^a (\xi) 
- \bar \theta\, C^a (\xi) +  \theta \, \bar\theta \, f^a (\xi),
\end{eqnarray}
where the coefficient of $\bar \theta$ is the BRST symmetry transformation: $s_b\, \xi^a = - \, C^a$ and that of $\theta$ is 
the anti-BRST symmetry transformation: $s_{ab}\, \xi^a = - \, \bar C^a$ where the fields $(\bar C^a)C^a$ are the (anti-)ghost
fields that have been taken into account for the anti-BRST and BRST symmetry transformations (which are 
the generalizations of the  {\it classical} infinitesimal diffeomorphism transformation parameters $\varepsilon^a (\xi)$
at the level of the {\it quantum} (anti-)BRST symmetry transformations within the framework of BRST formalism).  The secondary transformation parameter
$f^a (\xi)$ [which is the coefficient of $\theta\, \bar\theta$ in (95)] will be determined  by the requirement of the 
consistency condition which is nothing but, as we shall see later, the requirement of the  absolute anticommutativity 
for the (anti-)BRST transformations.

Within the framework of MBTSA to BRST formalism [7, 8], the target space coordinates 
 have to be generalized onto a (2, 2)-dimensional supermanifold in {\it two} steps as follows:
\begin{eqnarray}
X^\mu\, (\xi) & \longrightarrow &  X^\mu\, (\xi, \theta, \bar\theta), \nonumber\\
X^\mu \, (\xi, \theta, \bar\theta) \; &\longrightarrow & \;  \tilde{X}{^\mu} (\xi -  \theta\, \bar C 
- \bar \theta\, C +  \theta \, \bar\theta \, f, \; \theta, \;\bar\theta) \nonumber\\
& \equiv &  {\cal X}{^\mu} (\xi -  \theta\, \bar C  - \bar \theta\, C +  \theta \, \bar\theta \,f)  
+ \theta\, \bar R^\mu\, (\xi -  \theta\, \bar C  - \bar \theta\, C +  \theta \, \bar\theta \,f)\nonumber\\
& - & \bar\theta\,R^\mu\,  (\xi -  \theta\, \bar C  - \bar \theta\, C +  \theta \, \bar\theta \,f)
+\theta\,\bar\theta\,S^\mu \,  (\xi -  \theta\, \bar C  - \bar \theta\, C +  \theta \, \bar\theta \,f). 
\end{eqnarray} 
It is clear that ${\cal X}^\mu$ and $S^\mu$ are bosonic  secondary fields and the secondary fields
$(R^\mu, \, \bar R^\mu)$ are fermionic. 
We also point out that all the fields on the r.h.s. are function of the super 2D diffeomorphism transformation (95). 
Hence, we have to perform correctly  the mathematical trick of the Taylor expansion to obtain the following 
\begin{eqnarray}
&&\theta\,\bar\theta\,S^\mu \,  (\xi -  \theta\, \bar C  - \bar \theta\, C +  \theta \, \bar\theta \,f) = \theta\,\bar\theta\,S^\mu \,  (\xi),\nonumber\\
&&\bar\theta\,R^\mu\,  (\xi -  \theta\, \bar C  - \bar \theta\, C +  \theta \, \bar\theta \,f) = \bar\theta\, R^\mu (\xi) + \theta\,\bar\theta\,\bar C^a\partial_a
R^\mu (\xi), \nonumber\\
&&\theta\,\bar R^\mu\,  (\xi -  \theta\, \bar C  - \bar \theta\, C +  \theta \, \bar\theta \,f) = \theta\, R^\mu (\xi) - \theta\,\bar\theta\, C^a\partial_a
\bar R^\mu (\xi), \nonumber\\
&&{\cal X}^\mu \,(\xi -  \theta\, \bar C  - \bar \theta\, C +  \theta \, \bar\theta \,f) = X^\mu (\xi) - \theta\,\bar C^a\,\partial_a X^\mu - \bar\theta\,C^a\partial_a X^\mu \nonumber\\
&&~~~~~~~~~~~~~~~~~~~~~~~~~~~~~~~~~~+ \theta\,\bar\theta\, [f^a\,\partial_a X^\mu  - \bar C^a\, C^b \, \partial_a \partial_b X^\mu],
\end{eqnarray} 
where we have used the fermionic properties (i.e. $\theta^2 = 0, \, \bar\theta^2 = 0, \; \theta\, \bar\theta + \bar\theta\, \theta = 0$)
of the Grassmannian variables which anticommute with the secondary fields $R^\mu (\xi)$ and $\bar R^\mu (\xi)$. The above 
exercise has been performed because $X^\mu (\xi)$ is a Lorentz scalar w.r.t. the 2D 
infinitesimal diffeomorphism symmetry transformations on the 2D world-sheet (parameterized by $\xi^a$). 
On physical grounds, there have been two superspace generalizations [cf. Eq. (96)] because one has to incorporate the
diffeomorphism symmetry transformations over and above the generalization for the gauge symmetry transformations 
(see, e.g. [9-11])  
\begin{eqnarray}
X^\mu (\xi) \quad  \Longrightarrow \quad  X^\mu (\xi, \theta, \bar\theta) \quad  \Longrightarrow 
\quad \tilde X^\mu \, (\xi -  \theta\, \bar C  - \bar \theta\, C +  \theta \, \bar\theta \,f, \theta, \bar\theta).
\end{eqnarray} 
However, as we have pointed out,  the target space coordinates $X^\mu \, (\xi)$ 
are, precisely,  the Lorentz scalar fields w.r.t. the 
classical, infinitesimal and continuous diffeomorphism symmetry transformations 
[$\xi^a \rightarrow   \xi^a - \varepsilon^a (\xi)$] on the 2D world-sheet where $a = 0, 1$. In other words, the two 
step generalizations (98) will have {\it no} effect on these coordinates which are a scalar field w.r.t. the
2D world-sheet diffeomorphism symmetry transformations. At this stage, we exploit the ``horizontality 
condition"   that captures the above argument beautifully in the language of a simple  mathematical equation  as:
\begin{eqnarray}
 \tilde{X}{^\mu} (\xi -  \theta\, \bar C 
- \bar \theta\, C +  \theta \, \bar\theta \, f, \; \theta, \;\bar\theta)\; =  \;  X^\mu \, (\xi, \theta, \bar\theta) \equiv X^\mu (\xi).
\end{eqnarray} 
Collecting all the coefficients of $\theta, \, \bar \theta$ and $\theta\bar\theta$  in the 
expansions (97), we obtain  the following explicit expression  for the l.h.s. of the HC [cf. (99)], namely;
\begin{eqnarray}
&& \tilde X{^\mu} (\xi -  \theta\, \bar C 
- \bar \theta\, C +  \theta \, \bar\theta \, f, \; \theta, \;\bar\theta) = 
{X}^\mu  + \theta \, (\bar R^\mu - \bar C^a\, \partial_a\, {X}^\mu) + \bar \theta \, ( R^\mu - C^a\, \partial_a\, {X}^\mu) \nonumber\\
&&+ \theta \bar\theta \, (S^\mu + \bar C^a\, \partial_a\, R^\mu -  C^a\, \partial_a\, \bar R^\mu + f^a\, \partial_a\, X^\mu 
- \bar C^a\, C^b\, \partial_a\, \partial_b\, {X}^\mu) \equiv {X}^\mu (\xi).
\end{eqnarray} 
The restriction imposed by HC [cf. Eqs. (100), (98)], leads to the derivation of the explicit expressions  for the secondary fields 
$R^{\mu} (\xi), \, \bar R^{\mu} (\xi)$ and $S^\mu (\xi)$ as follows
\begin{eqnarray}
&&\bar R^\mu = \bar C^a\, \partial_a\, X^\mu, \qquad R^\mu = C^a\, \partial_a\, X^\mu, \nonumber\\
&&S^\mu =    C^a\, \partial_a\, \bar R^\mu  -\, \bar C^a\, \partial_a\, R^\mu - f^a\, \partial_a\, X^\mu 
+ \bar C^a\, C^b\, \partial_a\, \partial_b\, X^\mu.
\end{eqnarray} 
We would like to lay emphasis on the fact that the r.h.s. of equation [cf. Eqs. (100), (98)] is also equal to ${ X}^\mu (\xi, \, \theta, \, \bar\theta)$
where, in the first step, we have generalized the ordinary coordinates $X^\mu (\xi)$ to its counterpart ${ X}^\mu (\xi, \theta, \bar\theta)$
on the (2, 2)-dimensional supermanifold. In this step, it should be noted  that there is {\it no} diffeomorphism transformation
and it is just like the generalization in the context of gauge invariant theories (see, e.g. [9-11] for details) where a deep
 relationship was found by Bonora, {\it etal.}   (see, e.g. [9-11]) between the set of off-shell nilpotent (anti-)BRST symmetry transformations 
$s_{(a)b}$ and the translational generators $(\partial_\theta, \, \partial_{\bar\theta})$
along the Grassmannian direction of the ${\mathcal N} = 2$ generalized version of the supermanifold. In other words, it was shown 
by Bonora, Pasti and Tonin [9-11] that the following observation is true, namely; 
\begin{eqnarray}
{X}^\mu (\xi, \theta, \bar\theta) = X^\mu (\xi)  + \theta\, (s_{ab} X^\mu) + \bar\theta\, (s_b X^\mu) + \theta\, \bar\theta\, (s_b\, s_{ab}\, X^\mu),
\end{eqnarray} 
which shows that $s_b \longleftrightarrow \partial_{\bar\theta}\,|_{\theta = 0}$ and 
$ s_{ab} \longleftrightarrow \partial_{\theta}\,|_{\bar\theta = 0}.$
However, we  have to take another step in the context of diffeomorphism invariant theories so that the diffeomorphism generalization
($\xi^a \longrightarrow \xi^a - \theta\, \bar C - \bar\theta\, C + \theta\, \bar\theta\, f$) can be taken into account. Thus, we have
\begin{eqnarray}
 { X}^\mu \, (\xi, \theta, \bar\theta)  &\longrightarrow & \tilde{X}{^\mu} (\xi -  \theta\, \bar C 
- \bar \theta\, C +  \theta \, \bar\theta \, f, \; \theta, \;\bar\theta)\nonumber\\
& = &  {\cal X}{^\mu} (\xi -  \theta\, \bar C 
- \bar \theta\, C +  \theta \, \bar\theta \, f)  + \theta\, \bar R^\mu (\xi -  \theta\, \bar C 
- \bar \theta\, C +  \theta \, \bar\theta \, f)\nonumber\\
 & + & \bar\theta\,R^\mu \, (\xi -  \theta\, \bar C - \bar \theta\, C +  \theta \, \bar\theta \, f) + \theta\, \bar\theta\, S^\mu\, (\xi -  \theta\, \bar C 
- \bar \theta\, C +  \theta \, \bar\theta \, f).
\end{eqnarray} 
It is now obvious that, to obtain the 
off-shell nilpotent (anti-)BRST symmetry transformations $s_{ab} X^\mu, \, s_b X^\mu, \, s_b\, s_{ab} X^\mu$, we have to perform the appropriate 
Taylor expansions so that we can go from stage (103) to (102). This is what has been done {\it earlier} in our text.

To obtain the (anti-)BRST symmetry transformations, we have to find out the exact expressions for $R^\mu (\xi), \, \bar R^\mu (\xi)$
and $S^\mu (\xi)$ in terms of the  basic and auxiliary fields of the theory. We have already  obtained the precise expression for $R^\mu (\xi)$ 
and $\bar R_\mu (\xi)$ which imply the following (with standard inputs: $s_b \bar C^a  = i\,B^a$ and $s_{ab} C^a = i\,
 \bar B^a$):                                                     
\begin{eqnarray}
&&\bar R^\mu = \bar C^a\, \partial_a\, X^\mu = s_{ab} X^\mu,  \qquad R^\mu = C^a\, \partial_a\, X^\mu = s_b X^\mu, \nonumber\\
&&s_b s_{ab}X^\mu  = i\, B^a\, \partial_a X^\mu  - \bar C^a \, C^b\, \partial_a\partial_b\, X^\mu 
- \bar C^a\, (\partial_a C^b)\, (\partial_b X^\mu) \equiv S^\mu, \nonumber\\
&&-\,s_{ab} s_{b}X^\mu  = -\, i\, B^a\, \partial_a X^\mu  - \bar C^a \, C^b\, \partial_a\partial_b\, X^\mu
 + \bar C^a\, (\partial_a \bar C^b)\, (\partial_b X^\mu) \equiv S^\mu.
\end{eqnarray} 
In the above, the sanctity of the absolute anticommuting property (i.e. $\{ s_b, \, s_{ab}\} = 0)$ has been taken into account in the last two entries. 
On the other hand, the substitutions of $R^\mu (\xi)$ and $\bar R^\mu (\xi)$ into the expression for $S^\mu (\xi)$
leads to the following: 
\begin{eqnarray}
S^\mu  = [C^a\, \partial_a\bar C^b - \bar C^a\, \partial_a C^b - f^b]\, \partial_b X^\mu - \bar C^a
\ C^b\,\partial_a\, \partial_b\, X^b. 
\end{eqnarray} 
Comparison of (105), one-by-one, with the {\it two} such expressions (i.e. $S^\mu$)  in   (104) yields 
\begin{eqnarray}
f^a = -\,i\, B^a\,  + C^b\, \partial_b \bar C^a \equiv i\, \bar B^a - \bar C^b\, \partial_b\,C^a,
\end{eqnarray} 
which leads to (i) the precise determination of the secondary field $f^a (\xi)$, and (ii) the derivation of the CF-type restrictions:
$B^a + \bar B^a + i \; (\bar C^m \;\partial_m \;C^a + C^m \; \partial_m \;\bar C^a) = 0$ which is the 2D version of the D-dimensional 
CF-type restrictions that have been derived in our earlier work on the superfield approach to the D-dimensional diffeomorphism 
invariant theories (see, e.g. [7, 8] for details) where a bit of gravitation has been discussed, too.
The derivation of the (anti-)BRST invariant CF-type restrictions for the 2D diffeomorphism invariant model of the bosonic string theory [6], 
 to the best of our knowledge,  is a novel result. \\

\section{Conclusions}

In our present endeavor, we have focused on the one (0 + 1)-dimensional (1D) and two (1+1)-dimensional  (2D) diffeomorphism invariant (non-)interacting  
models of (i) the (non-) SUSY and (non-)relativistic particles, and (ii) the bosonic string of a specific kind (proposed by Kato and Ogawa [6])
and established that, within the framework of BRST and related supervariable/superfield approach, the Curci-Ferrari (CF) type restriction(s)
are {\it universal} in the sense that these are the special cases of the CF-type restrictions that have been derived for any arbitrary 
D-dimensional diffeomorphism invariant theory in our earlier works [7, 8]. 
The universal nature of CF-type restriction(s) is established using the theoretical strength of the MBTSA to BRST formalism [7, 8] where 
the {\it classical}, continuous and infinitesimal 1D and 2D diffeomorphism symmetries are elevated to the (anti-)BRST symmetry transformations.
This observation should be contrasted   with the (anti-)BRST symmetries and corresponding CF-type restriction(s) that exist in the 
cases of 1D (non-)SUSY, (non-)interacting  and (non-)relativistic particles when their {\it classical}, continuous 
and infinitesimal {\it gauge} symmetries 
are elevated to their {\it quantum} counterparts  (anti-)BRST symmetries. In the {\it latter} cases, 
the CF-type restriction(s) are {\it not} universal and they [even though
(anti-)BRST invariant] vary from one system to the other gauge invariant system(s).

Out of all the 1D systems (that have been discussed and described in our present endeavor), {\it only} one system 
(i.e. a free massive spinning relativistic particle) is endowed with a non-trivial CF-type restriction 
(i.e. $b + \bar b + 2\, \bar\beta\, \beta = 0$) in the context of quantum (anti-)BRST transformations
 corresponding to the {\it classical}, continuous and infinitesimal 
(super)gauge symmetry  transformations.  As a consequence, for the case of a massive spinning relativistic particle, 
we have a set of coupled (but equivalent) Lagrangians which respect the {\it quantum} (anti-)BRST symmetry  transformations\footnote{ In
our present endeavor, we have focused only on the existence of the nilpotent (anti-)BRST symmetries. However, we have been able to discuss
a few field-theoretic models of Hodge theory (within the framework of BRST formalism) where the nilpotent (anti-)co-BRST
symmetries also exist along with other bosonic continuous symmetries as well as a set of discrete symmetries
(see, e.g. [61] and references therein). The independent identities of the BRST, anti-BRST, co-BRST and anti-co-BRST
symmetries (and their corresponding conserved charges) become transparent in the contexts of such kinds of models
(see. e.g. [62] for details) where the physical realizations of the cohomological operators are provided. Furthermore, the
BRST and anti-BTST symmetries are needed {\it together} (in an independent fashion) for the derivation of the gauge-fixing and FP-ghost terms 
in the context of {\it all} the gauge/diffeomorphism invariant theories.} 
corresponding to $(i)$ the {\it classical}, continuous and infinitesimal (super)gauge symmetry  transformations, and 
$(ii)$ the classical, continuous and infinitesimal 1D diffeomorphism symmetry transformations. 
In the {\it rest} of the 1D systems, we have a {\it single} (anti-)BRST invariant Lagrangian corresponding to the 
classical, continuous and infinitesimal {\it gauge} symmetry transformations because of the fact that the CF-type restriction is {\it trivial}
(for all such 1D systems). However, we observe that there is {\it always} 
existence of a set of (anti-)BRST invariant coupled (but equivalent) Lagrangians for the cases where the classical, 
continuous and infinitesimal 1D and 2D diffeomorphism symmetry transformations are elevated to the level of {\it quantum} (anti-)BRST 
symmetry transformations. We also find that the gauge-fixing and FP-ghost terms are found to be the {\it same} for {\it all} the 1D 
diffeomorphism invariant (non-)interacting models of the (non-)SUSY and (non-)relativistic particles when they are considered within 
the framework of BRST formalism. This happens due to the {\it universal} form of the CF-type restriction
[i.e. $B + \bar B + i\,(\bar C\,\dot C - \dot{\bar C}\,C) = 0$] that exists for {\it all} the above 1D 
diffeomorphism (i.e. reparameterization)  invariant  tractable systems of physical interest.

In contrast to the 1D systems of  {\it all} kinds of  particles where the classical, continuous and infinitesimal gauge transformations 
(generated by the first-class constraints) exist (cf. Appendix D), there is {\it no} gauge transformation in the case of a 2D 
diffeomorphism invariant model of a bosonic string (that was proposed by Kato and Ogawa [6]). Thus, we have the existence of a set of 
appropriate Lagrangian densities that respect the (anti-)BRST symmetry transformations corresponding to the 
infinitesimal, continuous and {\it classical} version of the 2D diffeomorphism symmetry transformations. 
Using the theoretical potential of MBTSA, we have established that the CF-type restrictions, in the 
case of the 2D diffeomorphism invariant theory, is the generalization of the 1D version of the 
CF-type restriction that exists for the 1D  (non-)interacting  systems of the (non-)SUSY and (non-)relativistic point particles 
which have been discussed in our present endeavor. It is pertinent to point out that the 1D and 2D versions of the 
CF-type restriction(s) are the {\it special} cases of the D-dimensional 
CF-type restrictions that exist for the D-dimensional diffeomorphism invariant theory [7, 8].
This establishes the {\it universal} nature of the CF-type restriction(s).

We dwell a bit now  on the existence of the CF-type restriction(s) in the context of BRST formalism and its importance 
therein. We have seen in our present endeavor that the {\it classical} gauge symmetries are generated by the first-class constraints in the 
terminology of the Dirac prescription  for the classification of constraints [22, 23]. These gauge symmetries can be exploited within 
the framework of BRST formalism for the covariant canonical quantization (of the gauge theories). Thus, the most modern and 
fundamental definition of a {\it classical} gauge theory is  the observation that 
{\it this} theory  is always endowed with the first-class constraints which generate the 
{\it local} gauge transformations. In other words,  only {\it those} transformations  are called as the gauge 
transformations (cf. Appendix D) that owe their origin (see, e.g. [25, 26]) to the first-class constraints 
in Dirac's terminology of the classification scheme
of constraints (see, e.g. [22-26]). In exactly similar fashion, 
we state that when a gauge and/or diffeomorphism invariant theory is 
{\it quantized} (within the framework of BRST formalism), the characteristic feature of the BRST-quantized theory is the 
observation that the (non-)trivial CF-type restriction(s) exist 
in the context of the quantized gauge and/or diffeomorphism invariant theory. 
For the gauge theory, discussed within the framework of BRST formalism, the CF-type restriction(s) are {\it not} universal
and their form and number vary from a specific system to the other system(s) (see, e.g. [30-32] for details). However, for the 
D-dimensional diffeomorphism invariant theory, we have been able to show that the form of the CF-type restrictions is:
$B_\mu + \bar B_\mu  + i\, (\bar C^\rho \, \partial_\rho\, C_\mu +  C^\rho \, \partial_\rho\, \bar C_\mu) = 0$
(with $\mu, \, \rho = 0, 1, 2, ..., D-1$) which has been found to be universal (see, e.g. [30-34]).
Here the bosonic fields $(B_\mu, \bar B_\mu)$ are the Nakanishi-Lautrup type auxiliary vector fields and the fermionic 
(anti-)ghost vector fields are denoted by $(\bar C_\mu)C_\mu$. It is straightforward to note that, in our present endeavor,
we have obtained the 1D and 2D versions  of this {\it universal}  
 CF-type restrictions that exist in any arbitrary D-dimensional spacetime.

One of the hallmarks of the gravitational as well as the (super)string theories is the diffeomorphism invariance. 
Thus, our  present endeavor is a precursor to the BRST and related superfield approach to the diffeomorphism invariant models 
of physical interest in the physical four (3+1)-dimensions of spacetime. This is the future perspective and scope of our 
present review-cum-research article. We have plans to pursue {\it this} direction in our future endeavors. The  results  of 
our research works will be reported elsewhere. \\

\vskip 1cm

\noindent
{\bf Acknowledgments}\\

\noindent
All the authors gratefully acknowledge the contributions by Dr. Ankur Tripathi in some of the
works that have been reported in the present endeavor. They also thank him for many useful discussions on
the topics that have been covered in this review-cum-research article. Critical and constructive remarks by our 
esteemed {\it Reviewer} are thankfully acknowledged, too,  which have improved the quality, clarity and lucidity of our presentation.\\

\vskip 0.9cm

\noindent
{\bf Data Availability}\\

\noindent
No data were used to support this study.\\

\vskip 0.7cm

\noindent
{\bf Conflicts of Interest}\\

\noindent
The authors declare that there are no conflicts of interest.\\

\vskip 0.7cm

\begin{center}
{\bf Appendix A: On the Singular Nature of the Lagrangian for the Non-SUSY 1D Diffeomorphism Invariant  Scalar Relativistic Particle }\\
\end{center}

\vskip 0.5 cm 

\noindent 
To corroborate our assertion  that our starting Lagrangian $L_0  = m\, \sqrt {{\dot x}^\mu \, \dot x_\mu} \equiv
\sqrt {\dot x_0 ^2 - \dot x_i ^2}\; (i = 1, 2,...D - 1)$ for the free scalar 
relativistic particle is {\it singular}, we take up 
the three (2 + 1)-dimensional (3D) flat Minkowski spacetime manifold as the target space for which we have:  
$\alpha, \beta, \gamma...\mu, \nu, \lambda... = 0,1,2$ and $i, j, k... = 1, 2$.
In  this case, we have the following $3\times 3$  {\it symmetric} matrix (see, e.g. [24, 23] for details)
\begin{eqnarray*}
W_{\alpha\beta} \; = \;
\renewcommand\arraystretch{2}\large{
\begin{pmatrix}
W_{00} \;  & \; W_{01} \; & \; W_{02}
 \\ W_{10}  \;  & \;W_{11} \; & \; W_{12}
 \\  W_{20}  \;  & \;W_{21} \; & \; W_{22}
\end{pmatrix}
\quad \equiv \quad 
\renewcommand\arraystretch{2}
\begin{pmatrix}
\frac{{\partial^2 \,L_0}}{{\partial\, \dot  x^0 \, \partial\, \dot x^0}} \;  & \; \frac{{\partial^2 \,L_0}}{{\partial\, \dot  x^0 \, \partial\,\dot  x^1}}
\;  & \; \frac{{\partial^2 \,L_0}}{{\partial\, \dot  x^0 \, \partial\, \dot x^2}}
 \\ \frac{{\partial^2 \,L_0}}{{\partial\, \dot  x^1 \, \partial\, \dot x^0}} \;  & \; \frac{{\partial^2 \,L_0}}{{\partial\, \dot  x^1 \, \partial\,\dot  x^1}}
\;  & \; \frac{{\partial^2 \,L_0}}{{\partial\,  \dot x^1 \, \partial\, \dot x^2}}
\\ \frac{{\partial^2 \,L_0}}{{\partial\,  \dot x^2 \, \partial\, \dot x^0}} \;  & \; \frac{{\partial^2 \,L_0}}{{\partial\,  \dot x^2 \, \partial\,\dot  x^1}}
\;  & \; \frac{{\partial^2 \,L_0}}{{\partial\, \dot  x^2 \, \partial\, \dot x^2}}
\end{pmatrix},~~~~~~~~(A.1)}
\end{eqnarray*} 
where $x^\alpha = (x^0, x^1, x^2) \equiv (x_0, \underline {x}, y)$ and $x_\alpha = (x_0, x_1, x_2) \equiv (x_0, -\,\underline {x}, -\,y)$
so that we have: $ x_\mu \, x^\mu = x\cdot x = x^2 = x_0^2 - x_1 ^2 - x_2 ^2 \equiv x_0^2 - \underline {x} ^2 - y ^2 $. The
sign difference in $x^\mu$ and $x_\mu$ is due to our choice of the 3D metric $\eta_{\mu\nu}$ = diag $(+1, - 1, - 1) \equiv \eta^{\mu\nu}$
and the notation
 $x^1 = \underline x = -\,x_1$ has been adopted to differentiate between $x^2 = x_0 ^2 - \underline {x} ^2 - y ^2$ and $x_1 ^2 \equiv \underline {x} ^2$. 
As a result of this choice of notations, we have the explicit form of the above  $3\times 3$ matrix as: 
\begin{eqnarray*}
~~~~~~~~~~~~~~~ ~~~~~~~~~~~~~W_{\alpha\beta} \; = \;
\renewcommand\arraystretch{2}\large{
\begin{pmatrix}
\frac{{\partial^2 \,L_0}}{{\partial\,  \dot x_0 \, \partial\, \dot x_0}} \;  & \; \frac{{\partial^2 \,L_0}}{{\partial\,  \dot x_0 \, \partial\,  \dot {{\underline  x}}}}
\;  & \; \frac{{\partial^2 \,L_0}}{{\partial\,   \dot x_0\, \partial\, \dot y}}
 \\ \frac{{\partial^2 \,L_0}}{{\partial\, \dot {{\underline  x}}\, \partial\,\dot  x_0}} \;  & \; \frac{{\partial^2 \,L_0}}{{\partial\,  \dot {{\underline  x}} \, \partial\, \dot {{\underline  x}}}}
\;  & \; \frac{{\partial^2 \,L_0}}{{\partial\,  \dot {{\underline  x}} \, \partial\, \dot y}}
\\ \frac{{\partial^2 \,L_0}}{{\partial\,  \dot y \, \partial\, \dot x_0}} \;  & \; \frac{{\partial^2 \,L_0}}{{\partial\,  \dot y \, \partial\, \dot {{\underline  x}}}}
\;  & \; \frac{{\partial^2 \,L_0}}{{\partial\,  \dot y \, \partial\, \dot y}}
\end{pmatrix}.~~~~~~~~~~~~~~~~~~~~~~~~(A.2)}
\end{eqnarray*} 
The substitutions of {\it all} the entries into the above  $3\times 3$ matrix  (with proper considerations 
of the precise  notations) lead to the following  
\begin{eqnarray*}
~~~~~~~~~~~~~~~~W_{\alpha\beta} \; = \; \frac {m} {(\dot x^2)^{3/2}}
\renewcommand\arraystretch{2}\large{
\begin{pmatrix}
- \, (\dot {{\underline x}^2} + \dot y^2) \; & \; \dot x_0\, \dot {{\underline  x}} \; & \; \dot x_0\, \dot y
\\ \dot x_0\,\dot {{\underline  x}} \; & \; (\dot y^2 - \dot x_0^2)  \; & \; -\, \dot {{\underline  x}}\, \dot y
\\ \dot x_0\, \dot y \; & \; -\, \dot {{\underline  x}}\, \dot y \; & \; (\dot {{\underline  x}^2} - \dot x_0^2) 
\end{pmatrix},~~~~~~~~~~~(A.3)}
\end{eqnarray*}
where we have taken into account: $\dot x ^2 - {\dot x}_0 ^2 = -\, (\underline {\dot x}^2 + \dot y^2).$ 
Now, at this stage, it is straightforward to  note that we have the following expression for the determinant, namely; 
\begin{eqnarray*}
~~~~~~~~~~~~~~~~~~~ \mbox {det} \; W_{\alpha\beta}  = [m^3/(\dot x^2)^{9/2}]\, [\dot x_0^2\, \{ (\dot {{\underline  x}^2}+ \dot  y^2) - (\dot {{\underline x}^2} + \dot  y^2)\}] = 0.~~~~~~~~~~~~~~~~~~~~(A.4)
\end{eqnarray*}  
We end this Appendix with the {\it final} remark that we have shown that  det $W_{\alpha\beta}  = 0$ for the 
$(2\times 2)$ and ($3\times 3$) matrices (in our present endeavor)   for the $D = 2, 3$ dimensional Minkowskian flat spacetime 
manifolds that have been taken as the  target space, respectively. These observations strongly indicate that det $W_{\alpha\beta}  = 0$ for the
higher D-dimensional (i.e. $D = 4, 5, 6,...$) Minkowskian flat spacetime target spaces, too. However, one has to prove it by taking into account 
only {\it one} specific dimension  of the target space, at a time. \\

\vskip 1.2 cm

\begin{center}
{\bf Appendix B: On the Singular Nature of the Starting Lagrangian for the Free SUSY Massive Spinning  Relativistic Particle }\\
\end{center}

\vskip 1.2 cm 

\noindent
 We take into consideration the three (2 + 1)-dimensional (3D) flat Minkowskian spacetime 
manifold as the target space to show that, even in the 3D case, the determinant of the $3 \times 3$ symmetric 
matrix $W_{\alpha\beta}$ turns out to be zero. For this purpose, we focus on the diffeomorphism as well as gauge invariant starting Lagrangian $L_0$ of 
Eq. (31) and compute the following  $3 \times 3$ 
{\it symmetric} matrix which is denoted by $W_{\alpha\beta}$ (see, e.g. [24, 23] for details)
\begin{eqnarray*}
~~~~~~~~~~~~~~~~~~~W_{\alpha\beta} \; = \;
\renewcommand\arraystretch{2}\large{
\begin{pmatrix}
\frac{{\partial^2 \,L_0}}{{\partial\, \dot  x^0 \, \partial\, \dot x^0}} \;  & \; \frac{{\partial^2 \,L_0}}{{\partial\, \dot  x^0 \, \partial\,\dot  x^1}}
\;  & \; \frac{{\partial^2 \,L_0}}{{\partial\, \dot  x^0 \, \partial\, \dot x^2}}
 \\ \frac{{\partial^2 \,L_0}}{{\partial\, \dot  x^1 \, \partial\, \dot x^0}} \;  & \; \frac{{\partial^2 \,L_0}}{{\partial\, \dot  x^1 \, \partial\,\dot  x^1}}
\;  & \; \frac{{\partial^2 \,L_0}}{{\partial\,  \dot x^1 \, \partial\, \dot x^2}}
\\ \frac{{\partial^2 \,L_0}}{{\partial\,  \dot x^2 \, \partial\, \dot x^0}} \;  & \; \frac{{\partial^2 \,L_0}}{{\partial\,  \dot x^2 \, \partial\,\dot  x^1}}
\;  & \; \frac{{\partial^2 \,L_0}}{{\partial\, \dot  x^2 \, \partial\, \dot x^2}}
\end{pmatrix}}
\quad \equiv \quad \frac{{\partial^2 \,L_0}}{{\partial\, \dot  x^{\alpha} \, \partial\, \dot x^{\beta}}},~~~~~~~~~(B.1)
\end{eqnarray*} 
where we have to take the help of explicit form of $W_{\alpha \beta }$ like  Eq. (33) for computations.

At this stage, due to our choice of the metric tensor $\eta_{\mu\nu} = (+1, -\, 1, \, -\, 1) \equiv \eta^{\mu\nu}$, 
we have the explicit form of the notations:
$\dot X^\mu = (\dot X^0, \, \dot X^1, \, \dot X^2) = (\dot X_0, \, \dot {\underline X}, \, \dot Y)$ and 
$\dot X_\mu = (\dot X_0, \, \dot X_1, \, \dot X_2) = (\dot X_0, \, -\, \dot {\underline X}, \, -\,  \dot Y)$ so that the symbol 
$A = \dot X^\mu\, \dot X_\mu$ is expanded as: $A = \dot X^2 \equiv \dot X_0^2 - \dot X_1^2 - \dot X_2^2 = \dot X_0^2 - \dot {\underline X}^2 - \dot Y^2$.
These inputs lead to the following  $3\times 3$ {\it symmetric} matrix:
\begin{eqnarray*}
W_{\alpha\beta} \; = \;
\renewcommand\arraystretch{2}\large{
\begin{pmatrix}
W_{00} \;  & \; W_{01} \; & \; W_{02}
 \\ W_{10}  \;  & \;W_{11} \; & \; W_{12}
 \\  W_{20}  \;  & \;W_{21} \; & \; W_{22}
\end{pmatrix}}
\end{eqnarray*} 
\begin{eqnarray*}
\quad ~~~~~~~~~~~~~~~~ \equiv \quad \frac {m}{(A)^{3/2}}
\renewcommand\arraystretch{2}{
\begin{pmatrix}
-(\dot {\underline X}^2 + \dot Y^2) \;  & \; \dot X_0\, \dot {\underline X} \;  & \; \dot X_0\, \dot Y
 \\  \dot X_0\, \dot {\underline X}  \;  & \; (\dot Y^2 - \dot X_0^2)
\;  & \; -\, \dot {\underline X}\, \dot Y
\\  \dot X_0\, \dot Y \;  & \; -\,\dot {\underline X}\, \dot Y
\;  & \; (\dot {\underline X}^2 - \dot X_0 ^2)
\end{pmatrix}.~~~~~~~~~~~~~~~~(B.2)}
\end{eqnarray*} 
It should be noted that we have chosen the notations of $X^\mu$ and $X_\mu$ in the component forms such that there is a difference between 
$A = \dot X^\mu\, \dot X_\mu = \dot X^2$ and $ {\dot {\underline X}}^2 = {\dot {X}}_1 \, \dot {X}_1$. 
In other words, we have: $\dot X^2 = \dot X_0^2 - \dot X_1^2 - \dot X_2^2 = \dot X_0^2 - \dot {\underline X}^2 - \dot {Y}^2$
 and $\dot X_1^2 = {\dot {\underline X}}^2$, respectively.
The explicit computations of the above determinant leads to the following:
\begin{eqnarray*}
~~~~~~~~~~~~~~~~~~~~\mbox {det}\, W_{\alpha\beta} = \frac {m^3}{(A)^{9/2}}\, \Big[ \dot X_0 ^2\;
\{(\dot {\underline X}^2 + \dot {Y}^2)
 - (\dot {\underline X}^2 + \dot {Y}^2)\} \Big] = 0. ~~~~~~~~~~~~~~~~~~~~(B.3)
\end{eqnarray*} 
Thus, the starting Lagrangian $(L_0)$ of (31) is a {\it singular} Lagrangian. We have proven the {\it singular} 
nature of $(L_0)$ in two and three dimensional flat  Minkowskian  target spacetime manifold. These observations strongly indicate 
that det $W_{\alpha\beta}  = 0$ for the
higher D-dimensional (i.e. $D = 4, 5, 6,...$) flat Minkowskian target spaces, too. However, one has to prove it by taking into account 
only {\it one} specific dimension  of the {\it above} target spaces, at a time. It is also pretty obvious that  the starting Lagrangian $(L_0)$ [cf. Eq. (31)] 
is {\it singular} w.r.t. the fermionic ($\chi^2 = 0$) variable $\chi$ because it has no generalized velocity ($\dot \chi$) in it.
In fact, in the  whole theory (that is described by $L_0$), we have the existence of two constraints: 
$p^2 - m^2 \approx 0$ and $p_\mu\,\psi^\mu - m\, \psi_5 \approx 0$ which can be explicitly checked.\\

\begin{center}
{\bf Appendix C: On the Derivation of the (Anti-)BRST Symmetries for Other Variables of Our 
Interacting 1D Diffeomorphism Invariant  Theory}\\
\end{center}

\noindent
In this Appendix, we focus on the derivation  of the (anti-)BRST symmetries of the {\it rest} of the 
variables of the 1D diffeomorphism invariant {\it interacting} scalar relativistic particle {\it besides} the ones that have been derived
by using the theoretical strength of the MBTSA to BRST formalism [7, 8]. We exploit the requirements of 
(i) the off-shell nilpotency, (ii) the absolute anticommutativity, and (iii) the invariance 
of the CF-type restriction under the (anti-)BRST symmetry transformations for this purpose. In this regard, it  is 
pertinent to point out that the application of MBTSA  to BRST formalism leads to the derivation of the following 
(anti-)BRST symmetry transformations and CF-type restriction 
\begin{eqnarray*}
&&s_b x_\mu  = C\, \dot x_\mu, \qquad  s_b p_\mu  = C\, \dot p_\mu, \qquad  s_{ab} x_\mu  = \bar C\, \dot x_\mu, \nonumber\\   
~~~~~~~~~~~~~~~~~~~~~ && s_{ab} p_\mu 
 = \bar C\, \dot p_\mu, \qquad  B + \bar B + i\, (\bar C\, \dot C - \dot {\bar C}\, C) = 0,~~~~~~~~~~~~~~~~~~~~~~~~(C.1)
\end{eqnarray*} 
where we have {\it also}  used the standard (anti-)BRST symmetry transformations: $s_b \bar C = i\, B$ and $s_{ab} C = i\, \bar B$. 
It is straightforward to  note that the requirement of the off-shell nilpotency of {\it all} the above (anti-)BRST
symmetry transformations [i.e. $s_{(a)b}^2 = 0$] leads to:  
\[
s_b B = 0, \qquad s_{ab} \bar B = 0, \qquad s_b C  = C\, \dot C, \qquad s_{ab}\bar C  = \bar C\, \dot {\bar C}.
\eqno (C.2)\]
On the other hand, the requirement of the (anti-)BRST invariance of the CF-type restriction: 
[i.e. $B + \bar B + i\, (\bar C\, \dot C - \dot {\bar C}\, C) = 0$] leads to the following
\[
s_b \, \bar B = \dot {\bar B}\, C - \bar B\, \dot C, \qquad s_{ab} \, B = \dot B \, \bar C - B\, \dot {\bar C},
\eqno (C.3)
\]
provided we use the validity of the CF-type restriction on our theory. 
It is straightforward to check that the (anti-)BRST symmetry transformations [that have been 
derived in (C.2) and (C.3)] are off-shell nilpotent $[s_{(a)b}^2 = 0]$ of order two. 
Furthermore, it is interesting to point out that the requirement of the following absolute anticommutativity 
property on the (anti-)ghost variables, namely; 
\[
 \{s_b, \, s_{ab} \} \, \bar C = 0, \qquad \qquad \{s_b, \, s_{ab} \} \, C = 0, 
\eqno (C.4)
\]
{\it also} leads to the derivation of the (anti-)BRST symmetry symmetry transformations for the Nakanishi-Lautrup 
variables $(\bar B, \, B)$ quoted in (C.3). 
In addition, it can be checked that whatever (anti-)BRST symmetry transformations are derived, so far, are 
{\it proper} in the sense that they satisfy (i) the off-shell nilpotency property, and (ii) the absolute anticommutativity 
property. The invariance of the CF-type restriction, under the above (anti-)BRST symmetry transformations, 
is automatic and natural which demonstrates the {\it physical} nature of {\it this} restriction 
(i.e. CF-type restriction) on our theory.

We end this Appendix with the remark that the theoretical strength of the MBTSA to BRST formalism {\it plus} the requirements of the
off-shell nilpotency, absolute anticommutativity and invariance of the CF-type restrictions are {\it not} good enough to lead 
to the derivation of the (anti-)BRST symmetry transformations for the einbein variable $e\, (\tau)$ for 
the reparameterization invariant model of an {\it interacting} scalar relativistic particle with the electromagnetic field in the background. 
However, we know that the einbein variable $e(\tau)$ is like a ``gauge" variable under the classical gauge symmetry transformations
and transforms like a  ``scalar density"
under the {\it classical} diffeomorphism transformations. Hence, the following generalizations of the 
{\it classical} reparameterization symmetries  to their counterparts {\it quantum} (anti-)BRST symmetries, namely; 
\[
\delta_r\, e = \frac{d}{d\, \tau}\, (\varepsilon \, e) \; \; \longrightarrow \; \;  s_b\, e = \frac{d}{d\, \tau}\, (C \, e),
\]
\[
\delta_r\, e = \frac{d}{d\, \tau}\, (\varepsilon \, e) \; \; \longrightarrow \; \; s_{ab}\, e = \frac{d}{d\, \tau}\, (\bar C \, e),  
\eqno (C.5)
\]
satisfy all the requirements of the BRST formalism in the sense that the off-shell nilpotency and 
absolute anticommutativity properties are respected by the (anti-)BRST symmetry transformations quoted in (C.5)
provided we take into account the validity of the CF-type restriction
[i.e. $B + \bar B + i\, (\bar C\, \dot C + \dot {\bar C}\, C) = 0$]. \\

\begin{center}
{\bf Appendix D: On the First-Class Constraints and Gauge Transformations}\\
\end{center}

\noindent
The purpose of this Appendix is to show that the first-class constraints generate the symmetry transformations that are christened as
the {\it gauge} symmetry transformations. To corroborate this claim, we focus on the first-order Lagrangian $L_f$ [cf. Eq. (16)]
for the free scaler relativistic particle. It is an elementary exercise to to note that the conjugate momentum $\Pi_e$ (w.r.t.
the einbein variable $e$) is the primary constraint on this theory. In the language
of Dirac, this observation can be expressed as follows
\[
 \Pi_e = {\displaystyle \frac{\partial L_f}{\partial \dot e}} \approx 0, 
\eqno (D.1)
\]
where we have used the Dirac notation for the weakly zero (i.e. $\approx 0 $). Thus, we are allowed to take the first-order time 
derivative on the above constraint. We can  demand, as a consequence, the time-evolution invariance of the above primary constraint that leads 
to the derivation of the secondary constraint on our theory. The Hamiltonian formulation is the most suitable approach
to derive the set of secondary, tertiary, etc., constraints. However, for our simple system, Lagrangian formulation is good enough
to obtain the secondary constraint which emerges from the following EL-EoM, namely;
\[
{\displaystyle \frac{d} {d t} \Big (\frac{\partial L_f}{\partial \dot e} \Big) } =  {\displaystyle \frac{\partial L_f}{\partial  e}}
\qquad \Longrightarrow \qquad \dot \Pi_e = -\, \frac{1}{2}\, \big (p^2 - m^2 \big ) \approx 0. 
\eqno (D.2)
\]
In physical terms, the above equation implies the time-evolution invariance of the primary constraint that leads to the
derivation of the secondary constraint as: $-\, \frac{1}{2}\, (p^2 - m^2) \approx 0 $. A close look at the above 
primary and secondary constraints shows that {\it both} of them are expressed in terms of the canonical conjugate momenta
of our theory that is described by the first-order Lagrangian $L_f$. Hence, the commutator between the two is zero.
As a consequence, according to Dirac's prescription of the classification scheme of constraints [22,23], both 
the above constraints belong to the  first-class variety. There are {\it no} further constraints on our theory (which can be checked
by the Hamiltonian formulation).

The generator $G$, for the local, infinitesimal and continuous gauge symmetry transformations (17), can be written
in terms of the above first-class constraints by utilizing the standard theoretical technique (see, e.g. [26,25])
as follows: 
\[
G = \dot \xi \, \Pi_e + \frac{1}{2}\, \xi\, \big (p^2 - m^2 \big ). 
\eqno (D.3)
\]
It is straightforward to note that the following standard relationship exists between the continuous symmetry transformations ($\delta_g$)
and their generator ($G$), namely;
\[
\delta_g \Phi (\tau) = -\, i\, \big [\Phi (\tau), G \big ], 
\eqno (D.4)
\]
where $\Phi (\tau)$ is the generic variable. For our case of the scalar relativistic particle, the generic variable $\Phi$  is equal 
to the variables $x_\mu (\tau), p_\mu (\tau), e (\tau)$, the square bracket on the r.h.s. stands for the commutator
and $\delta_g$ represents the infinitesimal, local and continuous gauge symmetry transformations (17). Using the
canonical commutators: $[ e(\tau), \; \Pi_e (\tau) ] = i, \; [ x_\mu (\tau), \; p^\nu (\tau) ] = i\, \delta_\mu^\nu $
(and taking into account the {\it rest} of the commutators to be zero), it can be readily checked that the above relationship (D.4) 
produces the infinitesimal, local and continuous gauge symmetry transformations (17).

We conclude this short Appendix with a couple of final remarks. First, the 
existence of the first-class constraints (in the terminology of Dirac's prescription for the classification scheme
of constraints) on a theory is the sure signature for {\it this} particular theory to be  an example for the gauge theory. Second, 
the equations (D.3) and (D.4) are exactly like our earlier equations (54) and (55). However, the related canonical commutators are 
{\it quite} different (from our present case of a scalar relativistic particle).

\end{document}